\documentclass[varenna]{cimento}
\usepackage{amsmath,amssymb,amsfonts}
\usepackage{graphicx}
\usepackage[english]{babel}
\usepackage{physics}
\usepackage[dvipsnames]{xcolor}
\usepackage{aas_macros}

\usepackage{framed}
\usepackage{url}

\DeclareRobustCommand{\rchi}{{\mathpalette\irchi\relax}}
\newcommand{\irchi}[2]{\raisebox{\depth}{$#1\chi$}} 

\makeatletter
\newcommand{\oset}[3][1.4ex]{%
  \mathrel{\mathop{#3}\limits^{
    \vbox to#1{\kern-2\ex@
    \hbox{$\scriptstyle#2$}\vss}}}}
\makeatother

\colorlet{darkorange}{orange!90!black}
\colorlet{darkgreen}{green!60!black}

\colorlet{shadecolor}{orange!15}
\title{Phenomenological models of Cosmic Ray transport in Galaxies}
\author{Carmelo Evoli \atque Ulyana Dupletsa}
\institute{Gran Sasso Science Institute, Viale Francesco Crispi 7, 67100 L'Aquila, Italy \\ INFN-Laboratori Nazionali del Gran Sasso (LNGS), via G. Acitelli 22, 67100 Assergi (AQ), Italy}
\shortauthor{C.~Evoli \atque U.~Dupletsa}

\makeindex

\begin{document}

\maketitle

\begin{abstract}
When examining the abundance of elements in the placid interstellar medium, a deep hollow between helium and carbon becomes apparent.
Notably, the fragile light nuclei Lithium, Beryllium, and Boron (collectively known as LiBeB) are not formed, with the exception of Li7, during the process of Big Bang nucleosynthesis, nor do they arise as byproducts of stellar lifecycles.
In contrast to the majority of elements, these species owe their existence to the most energetic particles in the Universe. 

Cosmic rays, originating in the most powerful Milky Way's particle accelerators, reach the Earth after traversing tangled and lengthy paths spanning millions of years.
During their journey, these primary particles undergo transformations through collisions with interstellar matter. This process, known as \emph{spallation}, alters their composition and introduces secondary light elements in the cosmic-ray beam.

In light of this, the relatively large abundance of LiBeB in the cosmic radiation provides remarkable insights into the  mechanisms of particle acceleration, as well as the micro-physics of confinement within galactic magnetic fields.

These lecture notes are intended to equip readers with basic knowledge necessary for examining the chemical and isotopic composition, as well as the energy spectra, of cosmic rays, finally fostering a more profound comprehension of the complex high-energy astrophysical processes occurring within our Galaxy.
\end{abstract}


\section{Prelude}
\label{sec:prelude}

\begin{figure}[t]
\centering
\includegraphics[width=0.7\textwidth]{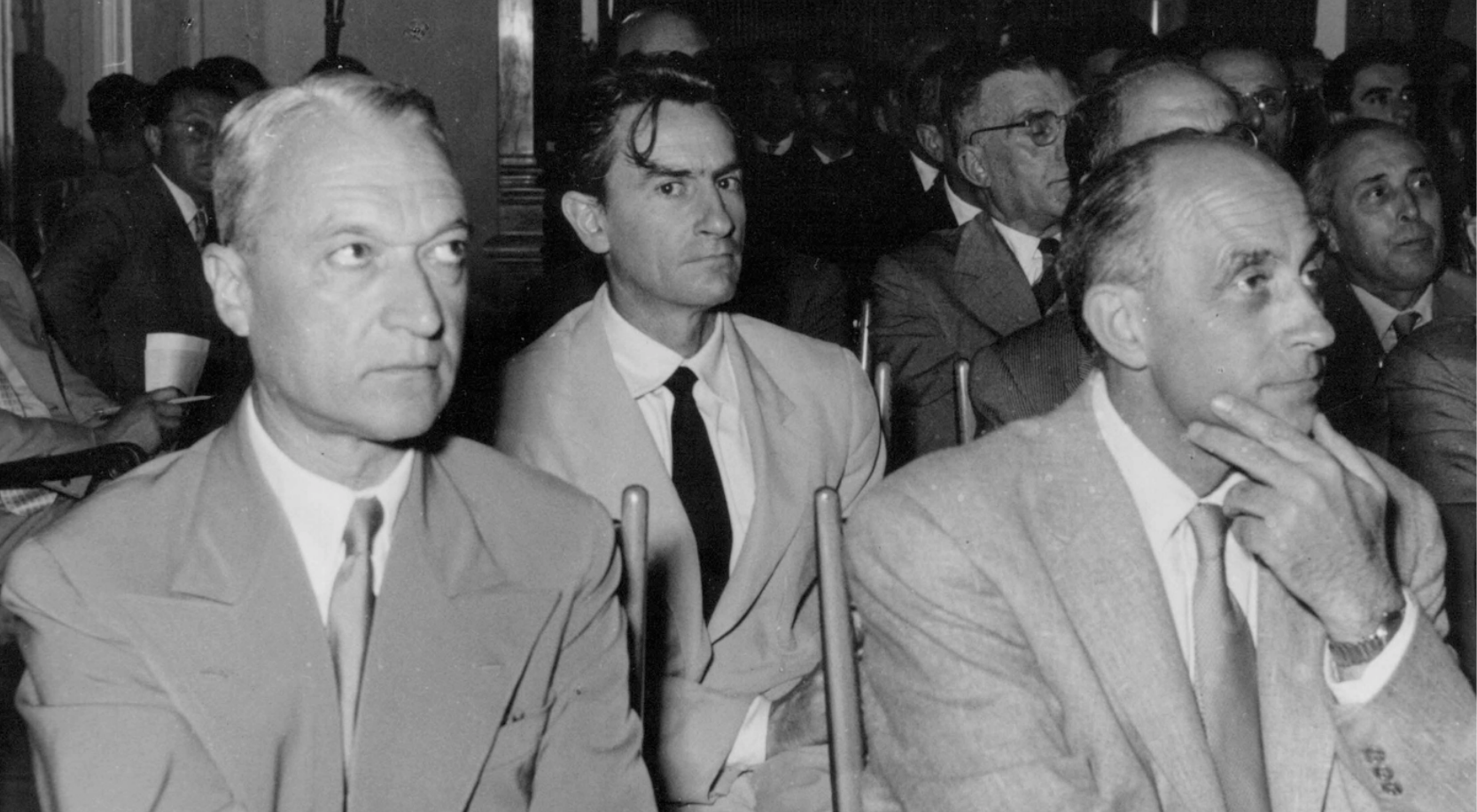}
\caption{Bruno Rossi (left) and Enrico Fermi (right) in Villa Monastero in 1954.}
\label{fig:rossifermi}
\end{figure}

In 1948, the Italian physicist Enrico Fermi attended a seminar in Chicago conducted by the Swedish physicist Hannes Alfvén, which focused on the motion of magnetic fields in dilute plasma and its potential connection to the origin of cosmic rays within the solar system (later published in~\cite{Alfven1949pr}).
Intrigued by Alfvén's arguments, Fermi approached him after the seminar and requested a more detailed explanation of ``magneto-hydrodynamic'' waves. According to Alfvén's recollections~\cite{Alfven1988amsci}, six years had passed since the publication of his initial paper on the subject~\cite{Alfven1942nat}, but only a few scientists acknowledged the existence of these waves, while most of his contemporaries dismissed the idea as foolish naivety.
After listening to Alfvén's explanation, for ``five or ten minutes'', Fermi responded, ``of course, such waves could exist!''. 
Fermi held such authority in the scientific community that, as Alfvén described it, ``if he said 'of course' today, every physicist said 'of course' tomorrow.''

Within a span of a few months, Fermi submitted a paper titled ``On the Origin of the Cosmic Radiation'' to the journal Physical Review, which was subsequently published in the April 15, 1949, issue, containing his prompt application of Alfvén's ideas~\cite{Fermi1949pr}.

In this seminal paper, Fermi utilized Alfvén's theory of the velocity of magnetoelastic wave propagation to propose the existence of magnetic fields in the rarefied interstellar medium. He made a surprisingly accurate prediction, given the time, that these fields should have a magnitude on the order of $5~\mu$G.

Furthermore, Fermi proposed that the intensity of these magnetic fields may vary in space, potentially being stronger within denser interstellar clouds. In this setup, charged particles would gain energy through head-on collisions with these clouds or lose energy during overtaking collisions. On average, head-on collisions were more probable, leading to a net energy gain for the particles.
As a consequence, ``the theory naturally yields an inverse power law for the spectral distribution of the cosmic rays,'' which remained unexplained at that time.

In hindsight, this mechanism proved to be too slow and inefficient to account for the observed energies of cosmic rays. Moreover, the power index depended on local details of the model, failing to generate a universal power law distribution. Nonetheless, this concept, known today as the \emph{second order Fermi mechanism}, became the foundation of many subsequent discussions on the subject, as it provided the minimum conditions necessary for the emergence of a power-law distribution in energy.

Of particular significance for our notes, Fermi's remarkable theory rested on the assumption that magnetic fields on a galactic scale existed within the interstellar medium, trapping cosmic rays. This proposal marked a pioneering hypothesis regarding the \emph{Galactic origin} of cosmic rays.

Fermi promptly presented his theory during the second edition of the International Cosmic Ray Conference in Como, Italy, in 1949, in a presentation titled ``Una teoria sull'accelerazione dei raggi cosmici''~\cite{Fermi1949icrc}.
He continued his work on this subject thereafter, and it is highly likely that he discussed his theory on galactic cosmic rays with Bruno Rossi during his last visit to Italy which took place during the second edition of the International School of Physics organized by the Italian Physical Society in Varenna in 1954 (figure~\ref{fig:rossifermi}).

With courage befitting the setting where Fermi delivered his last public seminar, the following lectures were given about 70 years later on the subsequent progresses of this topic.

\section{Introduction: the grammage pillar}
\label{sec:intro}

\begin{figure}
\centering
\includegraphics[width=0.65\textwidth]{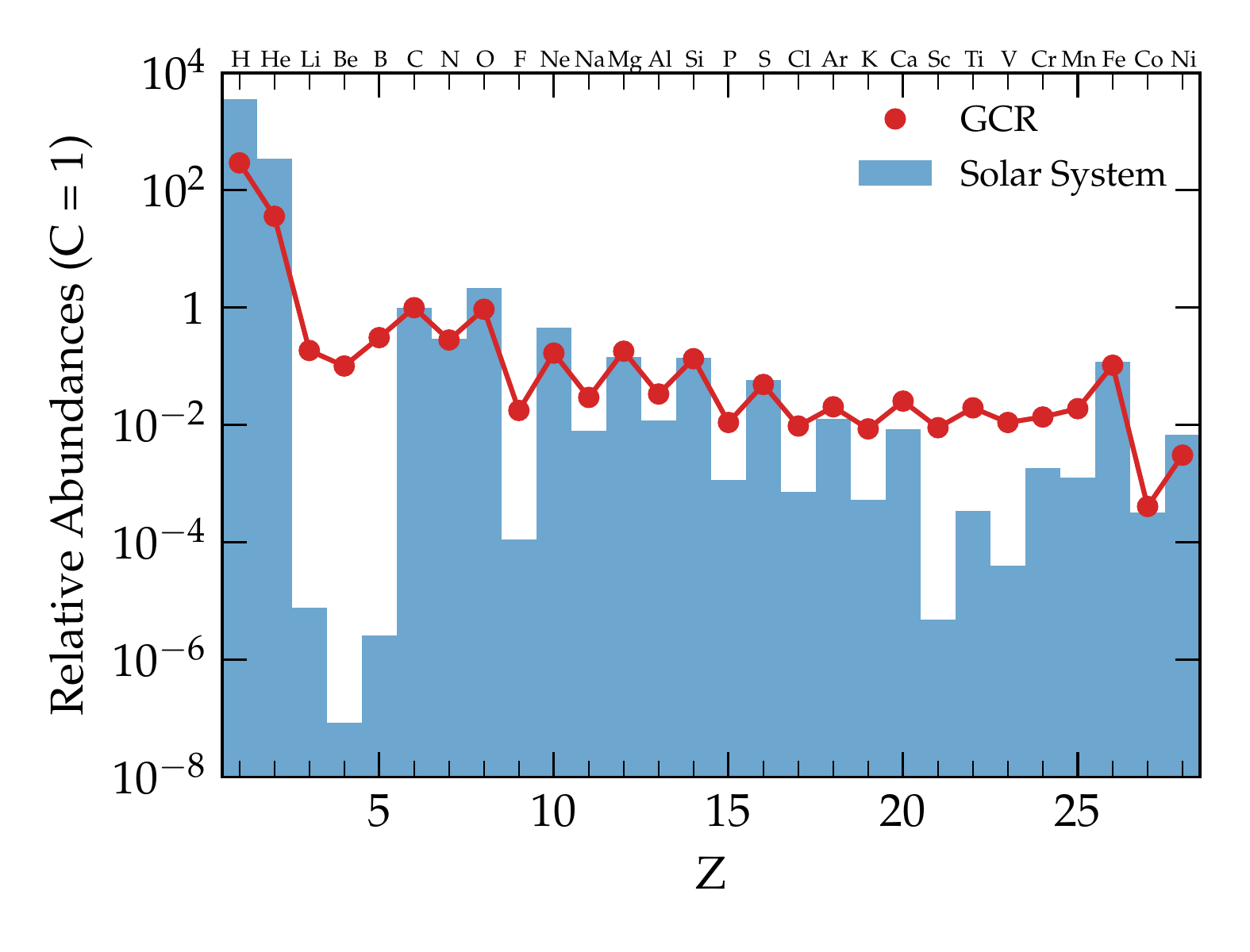}    
\caption{Abundances for Galactic cosmic rays (blue bars) and solar system elements (red circles) normalized to carbon atoms. Galactic cosmic ray abundances are a composite from~\cite{Young1981apj,AMS02results}. The solar system abundances are from~\cite{Looders2009}.}
\label{fig:composition}
\end{figure}

In figure~\ref{fig:composition}, we compare the isotope composition in the local cosmic ray (CR) flux against the composition in the ambient gas surrounding the solar system (detected, e.g., from CI-chondrites and solar photospheric measurements~\cite{Looders2009}). While the two compositions appear quite similar overall, suggesting that most CRs are accelerated from the average interstellar medium (ISM), intriguing differences can be observed.
Notably, elements like Lithium, Beryllium, and Boron in this plot closely resemble the abundance of Carbon or Oxygen in the CR flux, despite being negligibly present in the ISM on average. This feature is also confirmed for elements below Iron (Sc, Ti, and V), historically known as Sub-Iron elements, as well as for Nitrogen and Fluorine.

The striking separation between elements like Carbon and elements like Boron stands out. If we assume that the environment around the Sun is a typical representation of a Galactic star, reflecting a combination of stellar nucleosynthesis and pollution processes, and that some process within this environment is responsible for accelerating the surrounding elements, we would expect the relative ratio of CR abundances to match the interstellar one.
However, as our observations reveal, this is not the case. Thus, it indicates the presence of a second component, formed during propagation (hence termed \emph{secondary}), most likely resulting from the fragmentation of heavier CR nuclei into lighter ones during interactions with the ISM targets.

This secondary component provides relevant insights into the travel history of parent nuclei as they propagate through the ISM before reaching us.
As a matter of fact, it serves as one of the most compelling pieces of evidence supporting the idea of CR \emph{diffusive} propagation in our Galaxy.

To clarify this point, let's consider an extremely simplified scenario with only one primary species $n_p$ and one secondary species $n_s$. Their abundances can be described along the trajectory $s$ parameterized as \emph{grammage}, denoting the amount of material traversed by the particle during propagation\footnote{Analogous to the concept of ``column density'' in astrophysics}. 
In fact, the grammage, denoted by $\rchi$, is defined as the integral of the ISM density $\rho$ along $s$: $\rchi = \int \! ds \, \rho(s)$.

After an average interaction length $\lambda$, the CR nucleus (either primary, p, or secondary, s) undergoes an inelastic scattering with ISM targets, defined in terms of the inelastic cross-section $\sigma$ as $\lambda_{p(s)} = m / \sigma_{p(s)}$, where $m \sim 1.4~m_p$ is the mean target mass.

The evolution of the primary component involves only the disappearance of nuclei due to this effect. However, for the secondary component, we must consider production from the break-up of the primary component:
\begin{equation}
\begin{aligned}\frac{dn_p}{d\chi} & = - \frac{n_p}{\lambda_p} \\
\frac{dn_s}{d\chi} & = - \frac{n_s}{\lambda_s} + P_{p\rightarrow s} \frac{n_p}{\lambda_p}
\end{aligned}
\end{equation}
where $P_{p\rightarrow s}$ represents the spallation probability for producing $s$ from $p$ fragmentation.

By assuming that the system contains only primary nuclei for $\rchi = 0$ (initial condition), we can derive how the species ratio, $n_s / n_p$, depends on $\rchi$:
\begin{equation}
\frac{n_s}{n_p} = P_{p\rightarrow s} \frac{\lambda_s}{\lambda_s - \lambda_p} \left[ \exp\left( -\frac{\rchi}{\lambda_s} + \frac{\rchi}{\lambda_p} \right) - 1 \right]
\label{eq:chievolution}
\end{equation}

Laboratory experiments yield scattering lengths of about $\lambda_{\rm C} \sim 9.1$~g/cm$^2$ for primaries and $\lambda_{\rm B} \sim 10.4$~g/cm$^2$ for secondary nuclei. While the fragmentation probability is more uncertain, we assume a value of $P_{\textrm{C} \rightarrow \textrm{B}} \sim 0.25$~\cite{Evoli2019prd} for our purposes.

\begin{figure}
\centering
\includegraphics[width=0.6\textwidth]{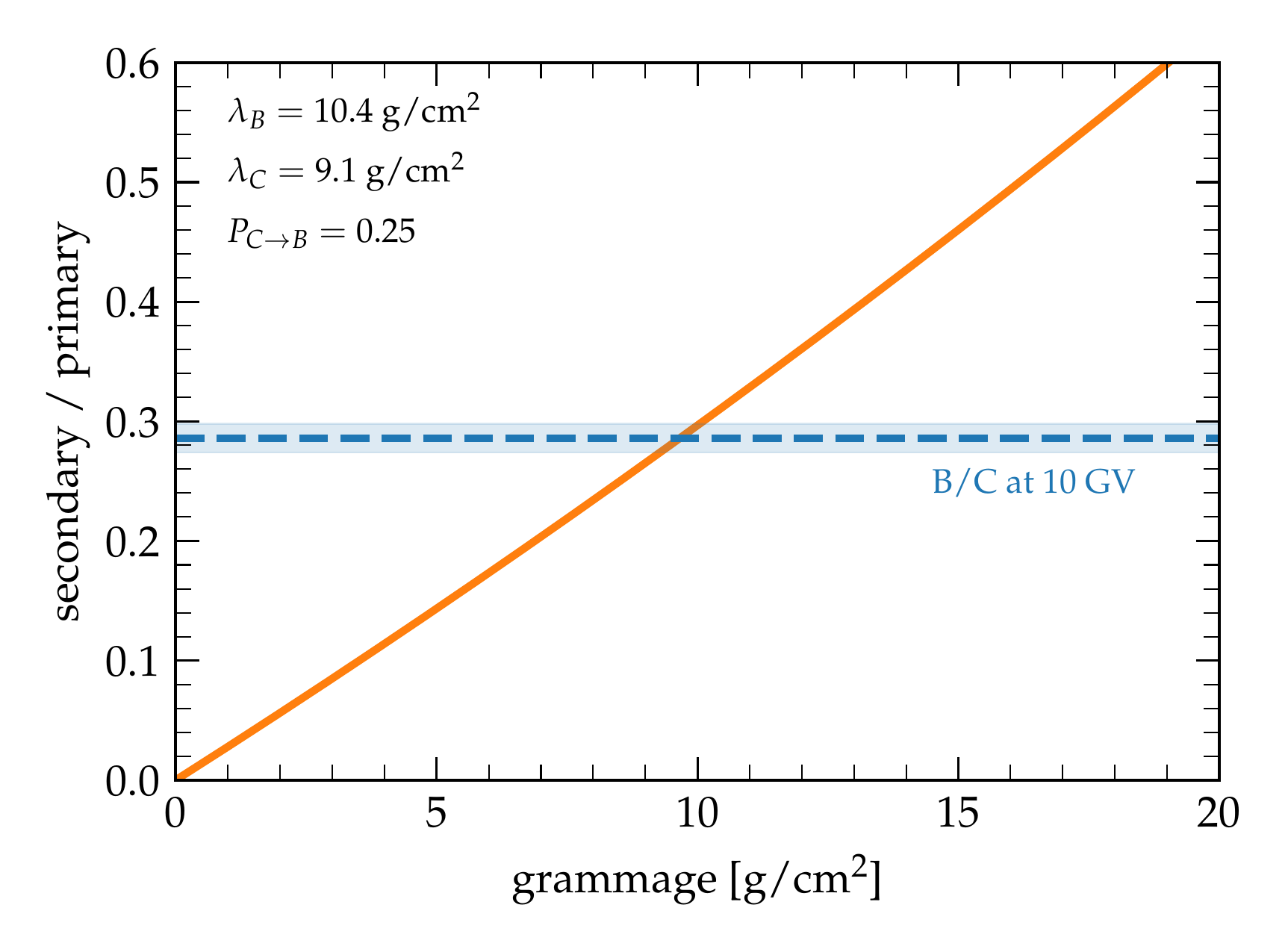}    
\caption{The secondary-over-primary ratio as a function of the grammage as given by equation~\eqref{eq:chievolution}. The value measured by AMS-02 at 10 GV is also shown with a dashed blue line~\cite{AMS02results}.}
\label{fig:grammage10gev}
\end{figure}

In figure~\ref{fig:grammage10gev}, we compare the ratio calculated using equation~\eqref{eq:chievolution} with the most recent measurements of the secondary-over-primary ratio by AMS-02~\cite{AMS02libeb} at a given rigidity $R = p/Z$. 
The comparison reveals the canonical \emph{$\mathcal O(10)$~grams per centimeters squared of traversed material for $\mathcal O(10)$ GV CRs}, a widely known fact in CR physics.

The gaseous component of the Milky Way is concentrated in a relatively thin disk with a half-thickness of approximately $h \sim 100$~pc. This disk has an average surface density of interstellar material, denoted as $\mu_d$, which is estimated to be around $\mu_d \sim 2.3 \times 10^{-3}$ g/cm$^{-2}$~\cite{Ferriere2001rmp}.

Given that the Galactic plane contains a significant portion of the GeV $\gamma$-ray emission, predominantly produced by the interactions of CR primaries with the interstellar material~\cite{Tibaldo2021universe}, it is reasonable to assume that the majority of the grammage accumulation occurs in this region.

However, if we consider a single crossing of the Galactic disk, the grammage accumulated is only roughly $\mu_d$. This value is clearly inconsistent with the estimation obtained from CR measurements. The discrepancy indicates that additional processes or regions contribute significantly to the total grammage encountered by CRs as they propagate through the Galaxy. 

The prevailing hypothesis is that particles cross the disk many times, and we can eventually estimate the minimum time that this primary component spends in the gas region in order to accumulate the required grammage as: 
\begin{equation}
\tau_{h} \gtrsim \frac{\chi_{\rm B/C}}{\mu_{\rm d}} \frac{2 h}{v} \sim 3 \times 10^6 \, \text{years}
\end{equation}

This timescale is clearly a factor $\frac{\chi_{\rm B/C}}{\mu_{\rm d}}$ larger than the typical time spent by a relativistic particle crossing straightly the disk which is $2h/c \sim 700$~years. 

The actual age of CRs in the Galaxy can be inferred by the relative abundances of radioactive nuclei. 
The measured abundances of CR clocks indicate a mean residence time for $\sim 10$~GV CR particles of about $\tau_{\rm H} \sim 6 \times 10^7$~yr\footnote{At this point, we kindly plead the reader's trust, as the justification for this value will be provided in \S\ref{sec:unstable}.}, which implies that CRs have to spend most of their time in low-density regions, with average density {$\bar n \lesssim (\mu_{\rm d} / 2h m) (\tau_{h}/\tau_{H}) \sim 0.1$~cm$^{-3}$}, not to overshoot the observed grammage.

To recap, the simple observation that the observed composition of CRs is different from that of solar elements in that rare solar-system nuclei such as B has provided compelling evidence that there is a process efficiently confining CRs within the Galaxy, enabling them to return multiple times from a low-density ``halo'' to the Galactic disk. 

The goal of these lecture notes is to develop a simplistic yet effective galactic model that can account for this fundamental observation and subsequently extract crucial physical quantities. 
In pursuit of this goal, we will discuss the foundational assumptions inherent to this class of models, while also outlining the consequential astrophysical implications that may capture also the interest of researchers in plasma and nuclear astrophysics.

These notes are intended for nonspecialists, and as such, the mathematical complexity of the model remains approachable. However, a broad spectrum of background knowledge is necessary, drawing from undergraduate courses in plasma physics, nuclear physics, and astrophysics.

While we leverage the notable advancements made in recent years within the realm of galactic CR physics, it's important to clarify that these notes are not intended as a comprehensive review of this subject.
To this end, we would like to refer to excellent recent monographs on this subject~\cite{Blasi2013aar,Zweibel2013pp,Grenier2015araa,Amato2018asr,Kachelriess2019ppnp,Gabici2019ijmpd}.

\section{Cosmic-ray protons in the Galaxy}
\label{sec:protons}

The local density of CRs, their energy spectrum, and their relative abundances provide us with the only direct information we can obtain about CRs. These measurements, along with estimates of the CR column density deduced from diffuse $\gamma$-rays~\cite{Tibaldo2021universe, Grenier2015araa}, form the foundation for any model aiming to describe CR transport.
In this section, our goal is to construct a basic model for CR propagation in the Galaxy that enables us to extract from these observables valuable information on the injection of CRs and the subsequent processes they undergo in the ISM.

As discussed in \S~\ref{sec:intro}, the escape time of CRs is too long to be compatible with straight propagation along the large-scale magnetic field. Consequently, CRs must be confined to the Galaxy for a significant period of time, primarily spending most of their time in low-density gas regions, such as the Galactic halo.
The existence of such an extended (larger than the gas disc) magnetized region is supported by observations of synchrotron emission from edge-on external galaxies~\cite{Beck2015aar}.

To account for these crucial aspects, virtually all CR transport models rely on the same basic ideas: high-energy particles are accelerated in sources located in a thin disc region of thickness $h \sim 100$ pc, following an injected spectrum proportional to $E^{-\gamma}$, with $\gamma \gtrsim 2$ as expected in the presence of diffusive shock acceleration mechanism (see D.~Caprioli's lecture notes in this volume).

\begin{figure}[t]
\centering
\includegraphics[width=0.6\textwidth]{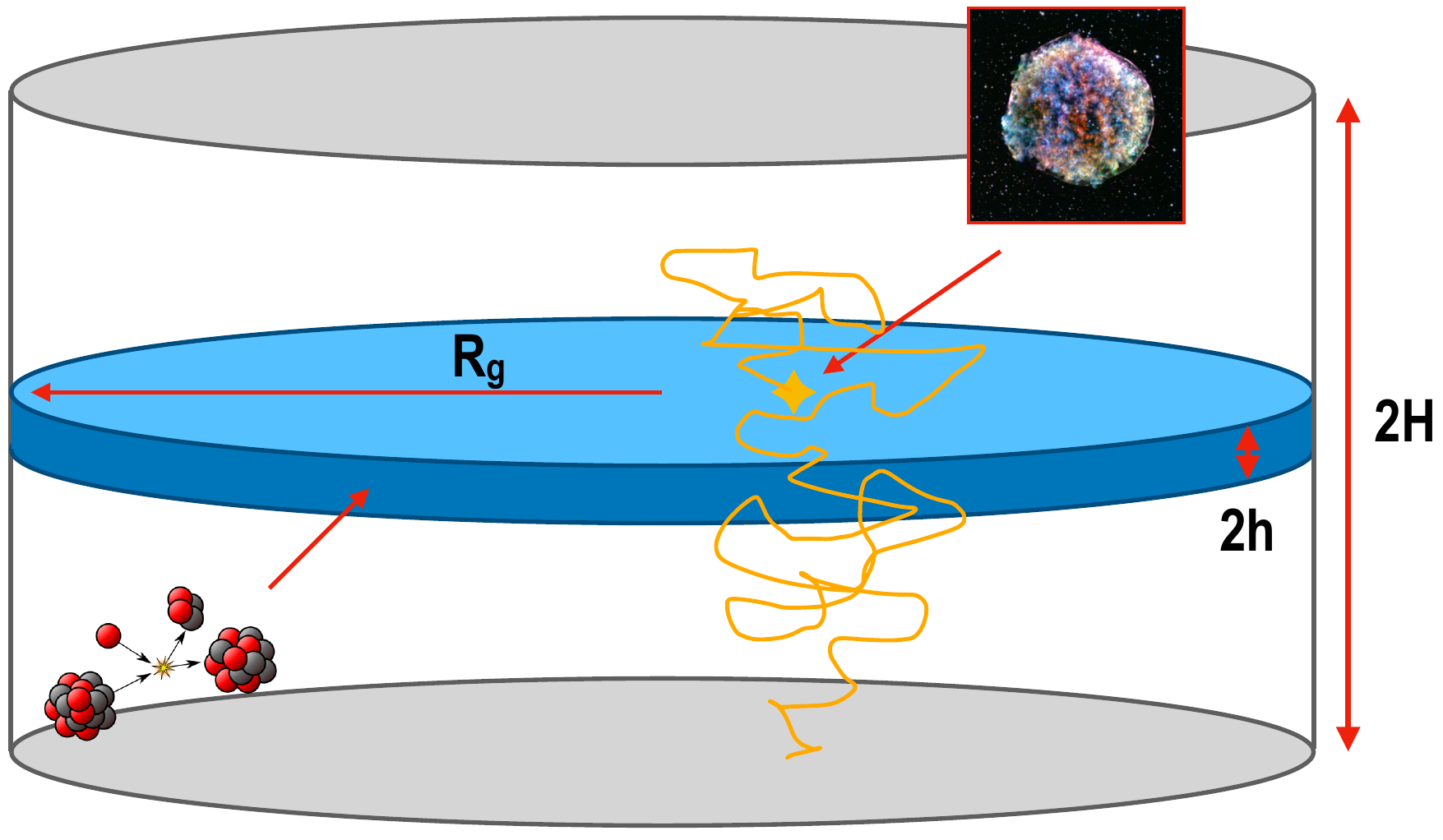}
\caption{Illustrative depiction of the Milky Way's diffusive halo from the perspective of a cosmic ray physicist. Primary sources are distributed only in the disk. Also secondary nuclei emerge from cosmic ray interactions within the gaseous disk.}
\label{fig:galaxy}
\end{figure}

After the injection, CRs propagate diffusively throughout the Galactic halo, which has a scale height $H$ of approximately several kiloparsecs with a diffusion coefficient $D$ proportional to $E^\delta$, where $\delta \sim 1/3 - 1/2$, and they escape freely at the boundaries. The escape is an essential, yet poorly understood, process to guarantee the stationarity of the problem, and is usually simplified by setting the CR density to zero at the boundary $H$ above and below the Galactic plane~\cite{Ginzburg1980apss} (see figure~\ref{fig:galaxy}).

Since the size of the halo is much smaller than the Galaxy's radius ($R_{\rm d} \gtrsim 10$ kpc), a one-dimensional model is sufficient to describe the escape of CRs along the z-direction.

The simplest representation of this framework is achieved by employing the diffusion equation in the form of Fick's law~\cite{Fick1855}. 
It is important to note that one could also utilize the more general Fokker-Planck equation (see P.~Blasi's lecture notes in this volume). 
Both Fick's law and the Fokker-Planck equation are considered as purely phenomenological equations, as they represent different choices for the flux in the fundamental continuity equation.

In general, we can write the continuity equation with a source term $Q(E, z)$ as
\begin{equation}
\frac{\partial n}{\partial t} + \frac{\partial J}{\partial z} = Q
\end{equation}
where $n(E, z, t)$ is the CR number density per unit energy and $J$ the corresponding flux along $z$.

According to Fick's law of diffusion, the diffusive flux is determined by the concentration gradient as $J = -D \nabla n$. In other words, a spatial gradient in the particle density will give rise to a current that transports particles from regions of high density to regions of low density, with the magnitude of the current being proportional to the diffusion coefficient $D$. The dimensions of $D$ are then \emph{area per unit time}.

The diffusion equation becomes
\begin{equation}
\frac{\partial n}{\partial t} = \frac{\partial}{\partial z} \left( D\frac{\partial n}{\partial z} \right) + Q
\end{equation}

For a spatially constant diffusion coefficient we can derive the associated Green’s function as 
\begin{equation}
\mathcal G(z,t) = \left(\frac{1}{{4 \pi D t}}\right)^{1/2} {\rm e}^{-\frac{z^2}{4Dt}}
\label{eq:green}
\end{equation}
which we interpret as the probability for finding a particle that is injected at the disc ($z = 0$) at a position $z$ after the time $t$.

The mean distance from the Galactic plane can be calculated from equation~\eqref{eq:green}:
\begin{equation}
\langle z \rangle = \left(\frac{1}{{4 \pi D t}}\right)^{1/2} \int dz z {\rm e}^{-\frac{z^2}{4Dt}} \simeq \sqrt{Dt}
\end{equation}

The characteristic time to reach a height $\langle z \rangle = H$ can then be defined as  $t_{\rm H} \simeq H^2 / D$, and the characteristic averaged velocity with which CRs escape from the Galaxy as
\begin{equation}
v_{\rm D} \sim \frac{H}{t_{\rm H}} \sim \frac{D}{H}
\end{equation}

It is important to acknowledge that in order to obtain the average distance from the galactic plane, we made the assumption that the diffusion coefficient remains spatially constant throughout both the halo and the disc. However, due to the variations in gas densities and magnetic field strengths between these regions, this assumption may not hold true.

By adopting the phenomenological assumption of diffusion as the primary transport process, we can construct the CR evolution equation by specifying the source term assuming Galactic Supernovae (SNe) as major contributors.

Since the galactic disc is extremely thin with respect to the halo, we can describe the spatial part of the injection term as a delta-function $\delta(z)$, and therefore
\begin{equation}
Q(E,z) = \frac{\xi_{\rm CR} E_{\rm SN} \mathcal R_{\rm SN} N(E)}{\pi R_{\rm d}^2} \delta(z) 
\end{equation}
where $E_{\rm SN} \simeq 10^{51}$~erg is the SN kinetic energy, $\xi_{\rm CR}$ is the fraction of this energy converted in CR acceleration, $\mathcal R_{\rm SN} \simeq 1 / 50$~yr$^{-1}$ is the rate of SNe in the Galaxy and $N(E)$ the spectrum of one SN. 

Considering that our Galaxy has an age of several billion years and observations of light elements suggest that CRs spend at most a hundred million years within our Galaxy, it can be concluded that the dynamical timescale of the Galaxy is significantly longer than the phenomena we are investigating. Therefore, we are justified in assuming stationarity, and the diffusion equation for protons $n_p$ becomes:
\begin{equation}
-\frac{\partial}{\partial z}\left[ D(E) \frac{\partial n_p}{\partial z}\right] = \frac{\xi_p E_{\rm SN} \mathcal R_{\rm SN}}{\pi R_{\rm d}^2} N(E) \delta(z)
\label{eq:protons}
\end{equation}

For $z \ne 0$, and imposing the boundary conditions $n_p(z = \pm H, E) = 0$, it gives the solution\footnote{The general solution to the differential equation can be written as $n_p(z)=A+Bz$:
\begin{itemize}
\item For $z>0$, we impose $n_p(H)=0$ which gives $n_p(z)=A\left(1-\frac{z}{H}\right)$
\item For $z<0$, we impose $n_{\rm p}(-H)=0$ which gives $n_{\rm p}(z)=A\left(1+\frac{z}{H}\right)$
\end{itemize}
Combining the two regions and imposing $n_p(0)=n_0(E)$, we obtain the solution valid in both regions as in equation~\eqref{eq:nzwithabs}.}:
\begin{equation}
D \frac{\partial n_p}{\partial z} = \text{Constant} \, \longrightarrow \, n_p(z) = n_0(E) \left( 1 - \frac{|z|}{H} \right)
\label{eq:nzwithabs}
\end{equation}

Thereby the diffusive flux is found to be constant in $z$, and we can compute it at the disc $z = 0$ as:
\begin{equation}
\left. D \frac{\partial n_p}{\partial z}\right|_{z=0^+} = - D \frac{n_{0}}{H}
\label{eq:flux}
\end{equation}

To find the density $n_0$ we integrate the diffusion equation around $z=0$:
\begin{equation}
\lim_{\epsilon\rightarrow0} \int_{\epsilon^-}^{\epsilon^+} \!\!\! dz 
\left\{ -\frac{\partial}{\partial z} \left[ D \frac{\partial n_p}{\partial z} \right] = Q(E,z) \right\} 
\end{equation}
which leads to
\begin{equation}
\left. -2D \frac{\partial n_p}{\partial z}\right|_{z=0^+} = \frac{\xi_p E_{\rm SN} \mathcal R_{\rm SN}}{\pi R_{\rm d}^2} N(E) 
\end{equation}
and using the equation for the flux derived in equation~\eqref{eq:flux}:
\begin{equation}
n_p(E) = \frac{\xi_p E_{\rm SN} \mathcal R_{\rm SN} N(E)}{2 \pi R_{\rm d}^2}  \frac{H}{D(E)} 
\label{eq:protonsimplesolution}
\end{equation}

\begin{figure}[t]
\centering
\includegraphics[width=0.6\textwidth]{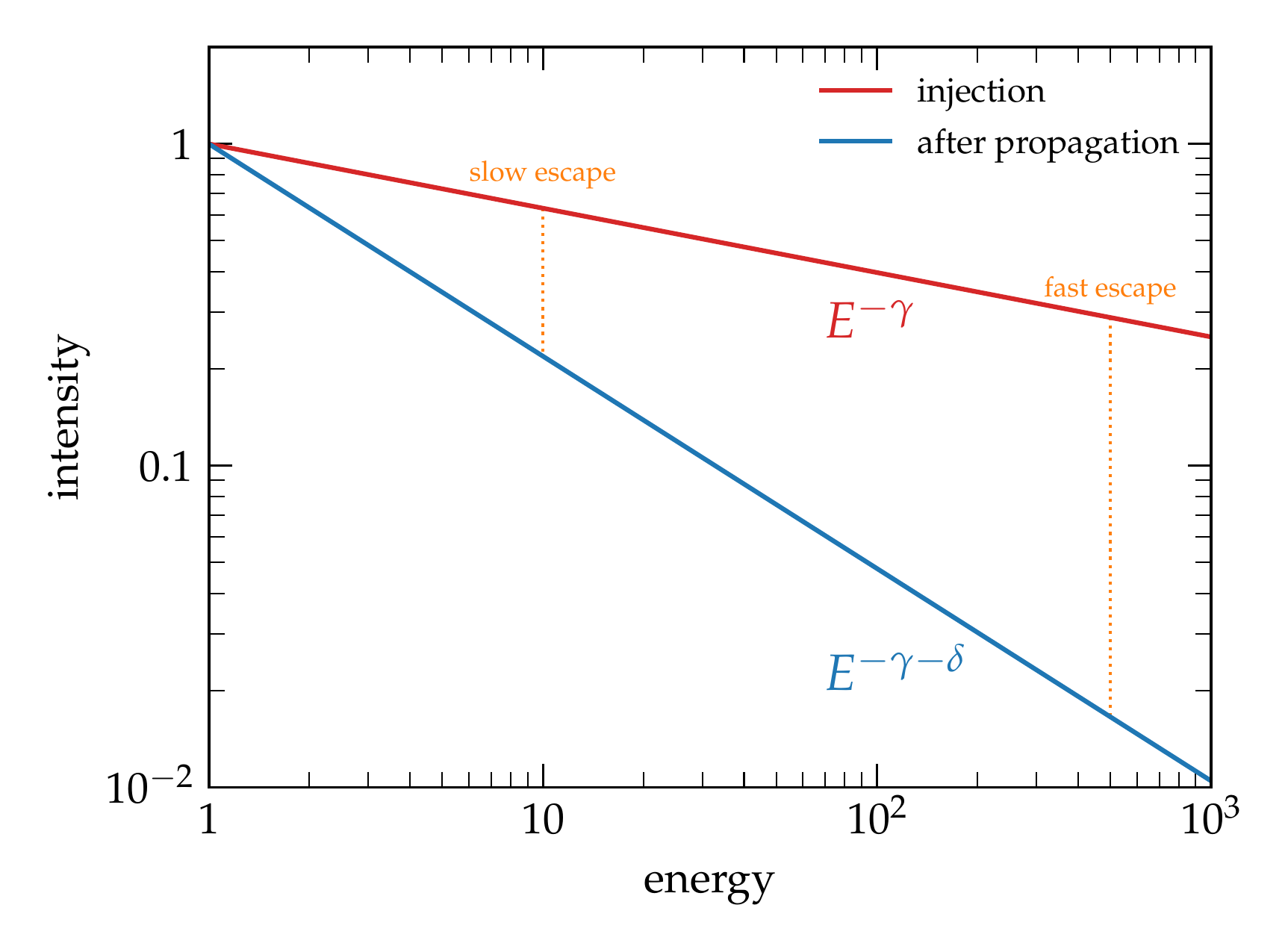}
\caption{Comparison between the \emph{injection spectrum} and the \emph{propagated spectrum} within an energy-dependent \emph{diffusion} toy-model. Higher-energy particles escape more swiftly, leading to a reduced equilibrium intensity. Consequently, the propagated spectrum consistently exhibits a \emph{steeper} profile than the injected counterpart.}
\label{fig:softening}
\end{figure}

As a result, we have obtained the CR spectrum as measured by an observer placed within the galactic disc.
It can be simply re-written as:
\begin{equation}
n_p(E) = \frac{Q_{\rm SN}(E) \tau_{\rm esc}(E)}{V_{\rm G}}
\end{equation}
where $Q_{\rm SN}(E) = \xi_p E_{\rm SN} \mathcal R_{\rm SN} N(E)$ is the number of particles injected per unit time by all supernovae, $\tau_{\rm esc} = H^2/D$ is the escape time, and $V_{\rm G} = 2\pi R_{\rm d}^2H$ is the total volume of the Galaxy.

In summary, since the diffusion coefficient increases with energy, following a power-law dependence as $D \propto E^{\delta}$, the escape time scales $\tau_{\rm esc} \propto E^{-\delta}$, and the spectrum we observe is always \emph{steeper} than the spectrum produced at the CR sources.

In point of fact, assuming that the sources inject in the ISM a power-law spectrum $E^{-\gamma}$, what we measure is a spectrum proportional to $E^{-\gamma-\delta}$ since high-energy particles spend less time in the Galaxy compared to lower-energy particles, resulting in a significant suppression of their density relative to the low-energy part of the spectrum.
This is depicted in figure~\ref{fig:softening}. 

It is important to note that since what we observe is a combination of the injection and transport processes, any deviation from a pure power-law in the observed spectrum of protons (or any primary species) cannot be easily disambiguate if due to propagation or acceleration effects. 

Finally, it is interesting to make some remarks on the assumptions we have made so far.  
The physics of CR transport is influenced not only by diffusion but also by boundary conditions\footnote{Further exploration of this subject is provided in the lecture notes by P.~Blasi within the same volume.}.
As discussed earlier, the condition of free escape, where $n(z = \pm H, E) = 0$, implies that the diffusion current does not depend on position.

At the boundary, conservation of flux requires:
\begin{equation}
\left. D \frac{\partial n}{\partial z} \right|_{z=H} = \frac{c}{2} n_{\rm out} 
\end{equation}
where it is assumed that particles outside the diffusing volume are streaming away approximately at the speed of light\footnote{The flux outgoing a semi-finite plane is $\Phi = \int_0^1 d\mu \mu v n \simeq n c \int_0^1 d\mu \mu$, where $\mu$ is the cosine of the incident angle.},~$c$.

On the other hand, we can express $D$ in terms of the mean free path $\lambda$ as $D = \frac{1}{3} c \lambda$\footnote{A plain  derivation of this expression is provided in Sec.~3.2 of~\cite{Kachelriess2008arxiv}}, yielding:
\begin{equation}
n_{\rm out} = \frac{2D}{cH} n_0 \sim \frac{\lambda(E)}{H} n_0 
\end{equation}

The condition of free-escape $n_{\rm out} \ll n_0$ is then satisfied as long as the path lenght $\lambda$ is much smaller than $H$, which remains true for $E \ll $~PeV~\cite{Blasi2019prl}. 

Despite the great importance of this assumption we do not have any knowledge on what determines the halo size or whether the halo size depends on energy or space (see however~\cite{Evoli2018prl,Dogiel2020apj} for an attempt to model the halo based on physical principles).

In the formulation of the transport equation for protons, we have initially disregarded the nuclear energy losses. For protons, the primary energy loss mechanism is the inelastic scattering with proton target leading to pion production. This process is characterized by a typical cross section of approximately $\sigma_{\rm pp}\sim 3\times 10^{-26}$ cm$^2$.

Nevertheless, when considering the timescale of this process in the ISM with a typical density of $n_{\rm d} \sim 1$ cm$^{-3}$, we find:
\begin{equation}
\tau_{\rm pp} \simeq \frac{1}{\bar{n}\sigma_{\rm pp}c} \simeq \, \text{Gyr}
\end{equation}

Here, $\bar{n} = n_{\rm d} h/H$ represents the average density within the diffusive volume.

From this calculation, we observe that the timescale for $pp$ scattering is significantly long, on the order of a Gigayear. Consequently, we can confidently assume that this process does not significantly alter the spectrum of propagated protons. 

However, despite being subdominant in terms of energy losses, the proton-proton interaction resulting in neutral pions that decay into two $\gamma$-rays is the leading mechanism for production of the spectacular diffuse emission observed along the Galactic Plane in Fermi-LAT data above $\sim 100$~MeV (see P.D.~Serpico's lecture notes in this volume).

On the other hand, the interaction cross-section becomes increasingly relevant when dealing with heavy nuclei. Therefore, in \S\ref{sec:nuclei}, we will discuss the incorporation of these effects in the transport of nuclei through the ISM.

\section{Primary nuclei}
\label{sec:nuclei}

Energy losses resulting from the fragmentation of nuclei in the ISM play a critical role in the propagation of CR nuclei across the Galaxy. It is worth noting that other loss mechanisms, such as ionization and Coulomb interactions, are only relevant for energies $T \lesssim 10$ GeV/n, and therefore, we will neglect them in the remainder of these lecture notes (see P.D.~Serpico's lecture notes in this volume).

When accounting for these losses, the transport equation for the phase-space distribution function\footnote{See appendix~\ref{app:intensity} for the formal definition of this quantity.} of a species $\alpha$ takes the following form:
\begin{equation}
-\frac{\partial}{\partial z} \left[ D_\alpha(p) \frac{\partial f_\alpha}{\partial z} \right] = 
Q_\alpha(p) \delta(z) - \frac{f_\alpha}{\tau_{\rm f, \alpha}} + \sum_{\alpha^\prime > \alpha}  \frac{f_{\alpha^\prime}}{\tau_{\rm f, \alpha\alpha^\prime}}
\end{equation}

In this equation, the source term $Q_\alpha(p)$ is now accompanied by a \emph{sink} term, which accounts for the spallation of CR nuclei. The rate of spallation is proportional to the spallation cross section $\sigma_\alpha$ and the density of the interstellar gas in the disc $n_{\rm d}$. Additionally, the last term on the right-hand side represents the source term due to the spallation of heavier nuclei of type $\alpha^\prime$ into nuclei of type $\alpha$. In that regard, the quantities $\tau_{\rm f, \alpha\alpha^\prime}$ contain the branching ratio of spallation of $\alpha^\prime$ into $\alpha$.

As the target gas for nuclear fragmentation is confined to the thin disk, we can explicitly define the timescale for inelastic losses as
\begin{equation}
\frac{1}{\tau_{\rm f, \alpha}} = 2 h n_{\rm d} \delta (z) c \sigma_\alpha
\label{eq:scalingcs}
\end{equation}
Additionally, we assume that spallation cross-sections are energy-independent. Interestingly, measurements of spallation cross-sections above a few GeV/n do not show any appreciable dependence on the projectile energy~\cite{Evoli2019prd}.

On the other hand, the cross sections of various elements increase with the atomic mass number, roughly following a geometric scaling law~\cite{Letaw1983apjs}:
\begin{equation}
\sigma_\alpha \simeq 45 \, \text{mb} \, A^{2/3}
\end{equation}

The source term is defined by assuming that particles are injected relativistically, $p \gtrsim p_{\rm min} = A m_p c$, with a power-law momentum distribution $p^{-\gamma}$:
\begin{equation}
Q_\alpha(p, z) = Q_{0,\alpha} \left(\frac{p}{p_{\rm min}}\right)^{-\gamma} \delta (z) 
\end{equation}

To determine the normalization factor, $Q_{0,\alpha}$, we impose that the total luminosity in CRs equals that released by Galactic SNe:
\begin{equation}
\xi_{\rm CR} E_{\rm SN} \mathcal R_{\rm SN} = \int \! dV \! \int \! d^3 \! p \, E(p) Q(z,p)
\end{equation}

This results in
\begin{equation}
Q_{0,\alpha} = \frac{(\gamma - 4) c^3 \xi_{\rm CR} E_{\rm SN} \mathcal R_{\rm SN}}{4 \pi^2 R_{\rm d}^2 E_{\rm min}^4}
\end{equation}
where $E_{\rm min} = p_{\rm min} c$.

The transport equation for a pure primary species, i.e., without a significant secondary contribution, can be written as
\begin{equation}
-\frac{\partial}{\partial z} \left[ D_\alpha(p) \frac{\partial f_\alpha}{\partial z} \right] = Q_{0,\alpha}(p)\delta(z) - 2 h n_{\rm d} \delta (z) c \sigma_\alpha f_\alpha(p)
\label{eq:primary}
\end{equation}

We notice that this equation can be solved in the same way and with the same boundary conditions at $z = \pm H$ as equation~\eqref{eq:protons}. For $z \neq 0$, it gives:
\begin{equation}
f_\alpha(z,p) = f_{0,\alpha}(p) \left( 1 - \frac{|z|}{H} \right)
\end{equation}

The expression of $f_{0,\alpha}(p)$ can be obtained by integration of equation~\eqref{eq:primary} around the disc
\begin{equation}
\left. -2 D_\alpha \frac{\partial f_\alpha}{\partial z} \right|_{0+} = Q_{0,\alpha}(p) - 2 h n_{\rm d} c \sigma_{\alpha} f_{0,\alpha}(p)
\end{equation}

Finally, we find:
\begin{equation}
f_{0,\alpha}(p) = \frac{Q_{0,\alpha} H}{2D_\alpha} \frac{1}{1 + n_{\rm d} \frac{h}{H} c \sigma_{\alpha} \frac{H^2}{D_\alpha}}
\end{equation}

Notice that $\bar n = n_{\rm d} h/H$ represents the \emph{average} density experienced by CRs during their Galactic propagation, while $H^2/D_\alpha$ denotes their diffusion time, and $1/(\bar n c \sigma_\alpha)$ represents the effective spallation time.

To express this in terms of the matter thickness traversed by CRs during their propagation, we introduce the concept of grammage, denoted as 
\begin{equation}
\rchi(p) = m_{\rm p} \left( n_{\rm d} \frac{h}{H} \right) c \left( \frac{H^2}{D_\alpha} \right) \sim m_p \bar n c \tau_{\rm esc}(p)
\label{eq:grammage}
\end{equation}

\begin{figure}[t]
\centering
\includegraphics[width=0.6\textwidth]{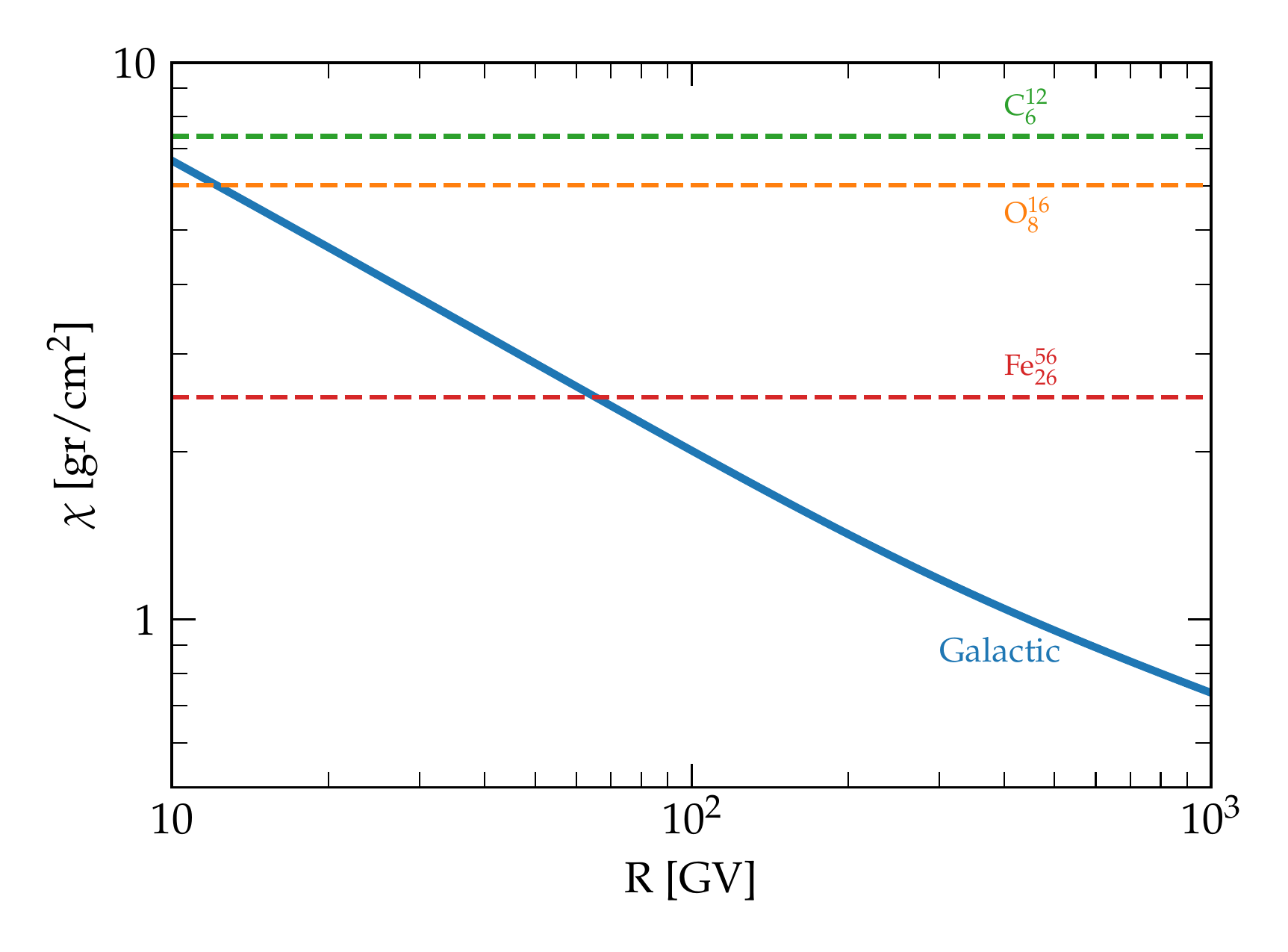}
\caption{The grammage corresponding to the best fit model described in~\cite{Schroer2021prd} as a function of rigidity is shown as a solid blue line. The dashed lines refers to carbon (green), oxygen (orange), and iron (red) inelastic critical grammage.}
\label{fig:grammage}
\end{figure}

Written in this form, the grammage serves to quantify the average density, $\bar n$, that particles traverse while moving across the Galaxy for a duration corresponding to the time it takes for particles to leave the Galaxy, $\tau_{\rm esc}$.

Furthermore, we define the critical grammage as 
\begin{equation}
\rchi_{\rm cr} = \frac{m_{\rm p}}{\sigma_\alpha} \simeq 40 \, A^{-2/3} \, \text{gr} \, \text{cm}^{-2}
\end{equation}

The critical grammage represents the threshold value where spallation becomes significant, requiring the grammage to be at least $\rchi_{\rm cr}$.

By using these definitions, the propagated spectrum of a species $\alpha$ can be expressed as:
\begin{equation}
f_{0,\alpha}(p) = \frac{Q_{0,\alpha}(p)}{2n_{\rm d}hm_{\rm p}c} \frac{1}{\frac{1}{\rchi(p)} + \frac{1}{\rchi_{\rm cr}}}
\label{eq:finalprimary}
\end{equation}

In the limit of strong and weak spallation, we obtain:
\begin{equation}
f_{0,\alpha}(p) = Q_{0,\alpha}(p) \times
\begin{cases}
\frac{1}{2n_{\rm d}hm_{\rm p}c}\rchi_{\rm cr} & \text{if } \, \rchi(p) \gg \rchi_{\rm cr} \\
\frac{H}{2D(p)} & \text{if } \, \rchi(p) \ll \rchi_{\rm cr}
\end{cases}
\end{equation}

Hence, when $\chi(p)\gg \chi_{\rm cr}$, spallation dominates, the spectrum at the disc will exhibit a similar slope to the injection spectrum. Conversely, in the limit of weak spallation, the diffusion-dominated solution is recovered, resulting in an observed spectrum that is \emph{steeper} than the source spectrum due to the momentum dependence of the diffusion coefficient.

At low energies, spallation may dominate because $D(p)$ is an increasing function of momentum, and spallation cross sections are independent of energy. For heavy nuclei, the effects become particularly relevant at higher energies due to the scaling with mass in equation~\eqref{eq:scalingcs}. In figure~\ref{fig:grammage} the critical grammage for different nuclei is compared with the Galactic value obtained by several recent analysis~\cite{Evoli2019prd,Weinrich2020aa}. For Iron, one of the heaviest species present in the Galactic radiation, the critical grammage becomes ruling for rigidities\footnote{We remind that rigidity is defined as momentum over charge, $p/Z$.} below $\lesssim 60$~GV~\cite{Schroer2021prd}.

Equation~\eqref{eq:finalprimary} highlights that if we could somehow measure the grammage, we would be able to retrieve the acceleration efficiency $\xi_{\rm CR}$ from the spectrum observed at Earth. This parameter is of crucial importance for the development of any acceleration paradigm.

However, obtaining this crucial information relies on a different observable, which will be discussed in detail in the next section.

\section{The secondary over primary ratio}
\label{sec:secondaryoverprimary}

As mentioned in the Introduction, light and fragile elements such as lithium, beryllium, and boron are mainly synthesized through the collisions of galactic CRs with the interstellar gas in the Galaxy. Here we investigate the production of these elements, and how they constrain CR models for Galactic propagation. 
For the sake of simplicity, we focus on a case with only carbon as the primary species ($\alpha' = \text{C}$), whereas boron is almost exclusively created in secondary processes ($\alpha = \text{B}$).

The solution for carbon has been obtained in the previous section and is given by:
\begin{equation}
f_{0,\rm C}(p) = \frac{Q_{0,\rm C}}{2n_{\rm d}h m_{\rm p}c} \frac{1}{\frac{1}{\rchi_{\rm C}(p)} + \frac{1}{\rchi_{\rm cr, C}}}
\end{equation}

Since boron is a pure secondary species, the solution of its transport equation takes the same form as the previous equation, with the injection rate proportional to the equilibrium solution for carbon:
\begin{equation}
4 \pi Q_{\rm B}(p) p^2 dp = 2 h n_{\rm d} c \sigma_{\rm C \rightarrow B} 4 \pi \delta(z) f_{\rm C}(p^\prime) p^{\prime \, 2} dp^\prime
\end{equation}

Here, $A$ is the atomic number of carbon, and the Jacobian $\frac{dp'}{dp} = \frac{A}{A-1}$ is introduced to ensure conservation of energy per nucleon.

The spectrum of boron in the disc is then given by:
\begin{equation}
f_{0,\rm B}(p) 
= \frac{\sigma_{\rm C \rightarrow B}}{m_{\rm p}}  \frac{1}{\frac{1}{\rchi_{\rm B}(p)} + \frac{1}{\rchi_{\rm cr, B}}} f_{0, \rm C}\left(\frac{A}{A-1}p\right) \left(\frac{A}{A-1} \right)^3
\end{equation}

This leads to the B/C ratio as:
\begin{equation}
\frac{\rm B}{\rm C} \simeq \frac{1}{\rchi_{\rm cr, C\rightarrow B}} \left(\frac{1}{\rchi(p)} + \frac{1}{\rchi_{\rm cr, B}}\right)^{-1}
\label{eq:chibc}
\end{equation}

It is important to note that the boron-to-carbon ratio is independent of the primary source and is solely a function of the grammage and relevant cross-sections.

Once again, considering the two limits of strong and weak spallation, we find the following. 
In the case of strong spallation:
\begin{equation}
\frac{\rm B}{\rm C} \oset{\rchi \gg \rchi_{\rm cr}}{\, \longrightarrow \,} \frac{\rchi_{\rm cr, B}}{\rchi_{\rm cr, C\rightarrow B}} 
= \frac{\sigma_{\rm C \rightarrow B}}{\sigma_{\rm B}} 
\simeq 0.3
\end{equation}

This results in a constant B/C ratio as a function of energy, with the value determined solely by the cross-sections~\cite{Evoli2019prd}.

In the opposite limit:
\begin{equation}
\frac{\rm B}{\rm C} \oset{\rchi \ll \rchi_{\rm cr}}{\, \longrightarrow \,} \frac{\rchi(p)}{\rchi_{\rm cr, C\rightarrow B}} \propto \frac{1}{D(p)}
\label{eq:bchene}
\end{equation}

Hence, in this limit, the B/C ratio is expected to decrease with increasing energy. 
Moreover, by measuring the B/C ratio as a function of energy, we can estimate the energy dependence of the grammage and, consequently, of the diffusion coefficient of Galactic CRs.

\begin{figure}
\centering
\includegraphics[width=0.6\textwidth]{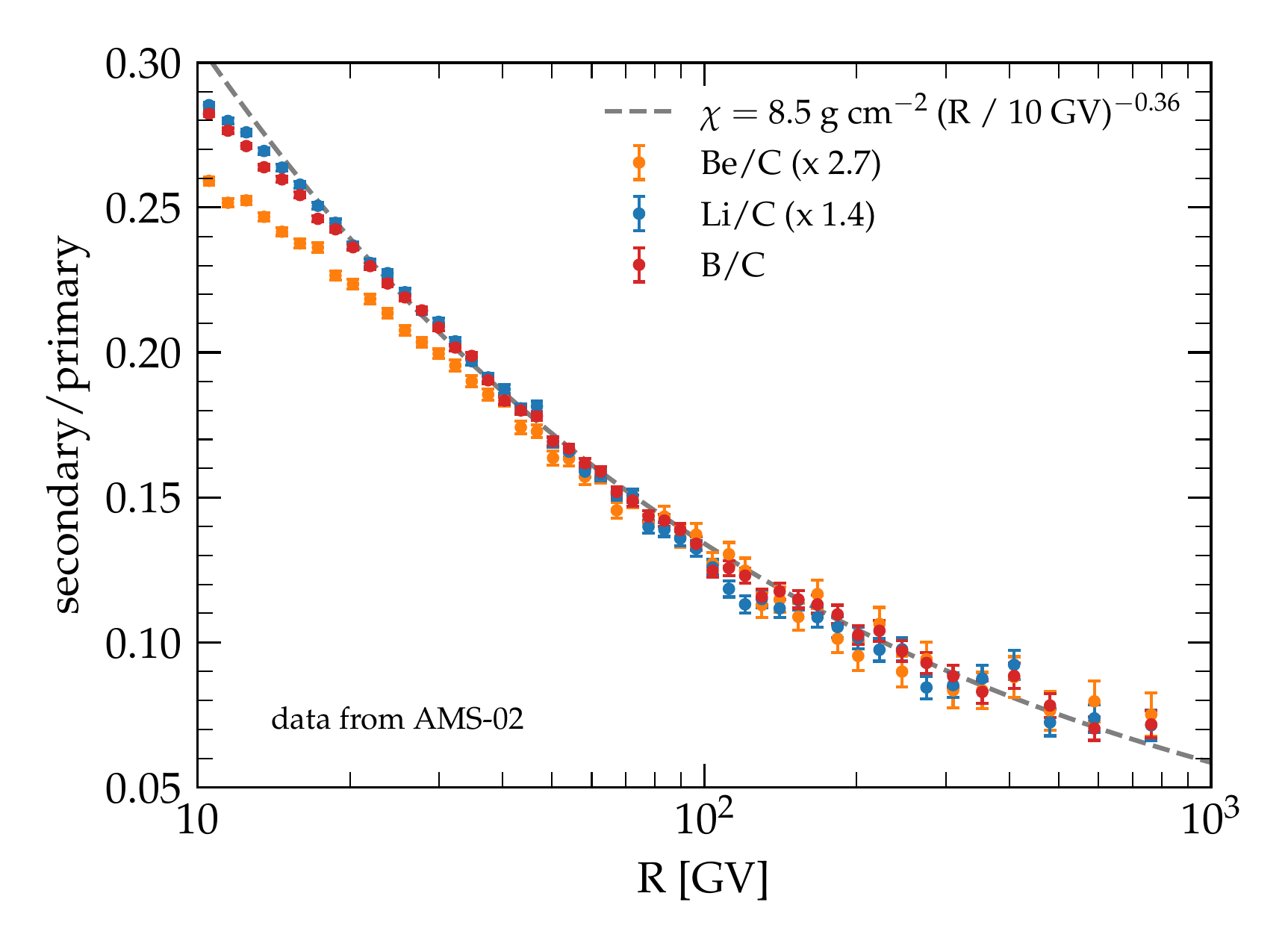}
\caption{Secondary-to-primary ratios as function of rigidity measured by the AMS-02 experiment~\cite{AMS02libeb}. The power-law fit at high energies is also shown.}
\label{fig:bcams02}
\end{figure}

It is difficult to overemphasize the significance of this result. 
The recent AMS-02 data on secondary-to-primary ratios (see figure~\ref{fig:bcams02}) clearly demonstrate that the B/C ratio decreases with increasing rigidity above $\sim$10 GV. 
This indicates that we are operating in the regime of weak spallation, allowing us to infer the Galactic grammage from this observable.
A quick fit to data shown in figure~\ref{fig:bcams02}, using $\sigma_{\rm C \rightarrow B} \simeq 60$~mb, indicates that the grammage is of the order of $8.5$~gr cm$^{-2}$ at $\simeq 10$~GV, exhibiting a scaling behavior approximately proportional to the rigidity raised to the power of $\simeq -1/3$. 

\begin{figure}
\centering
\includegraphics[width=0.6\textwidth]{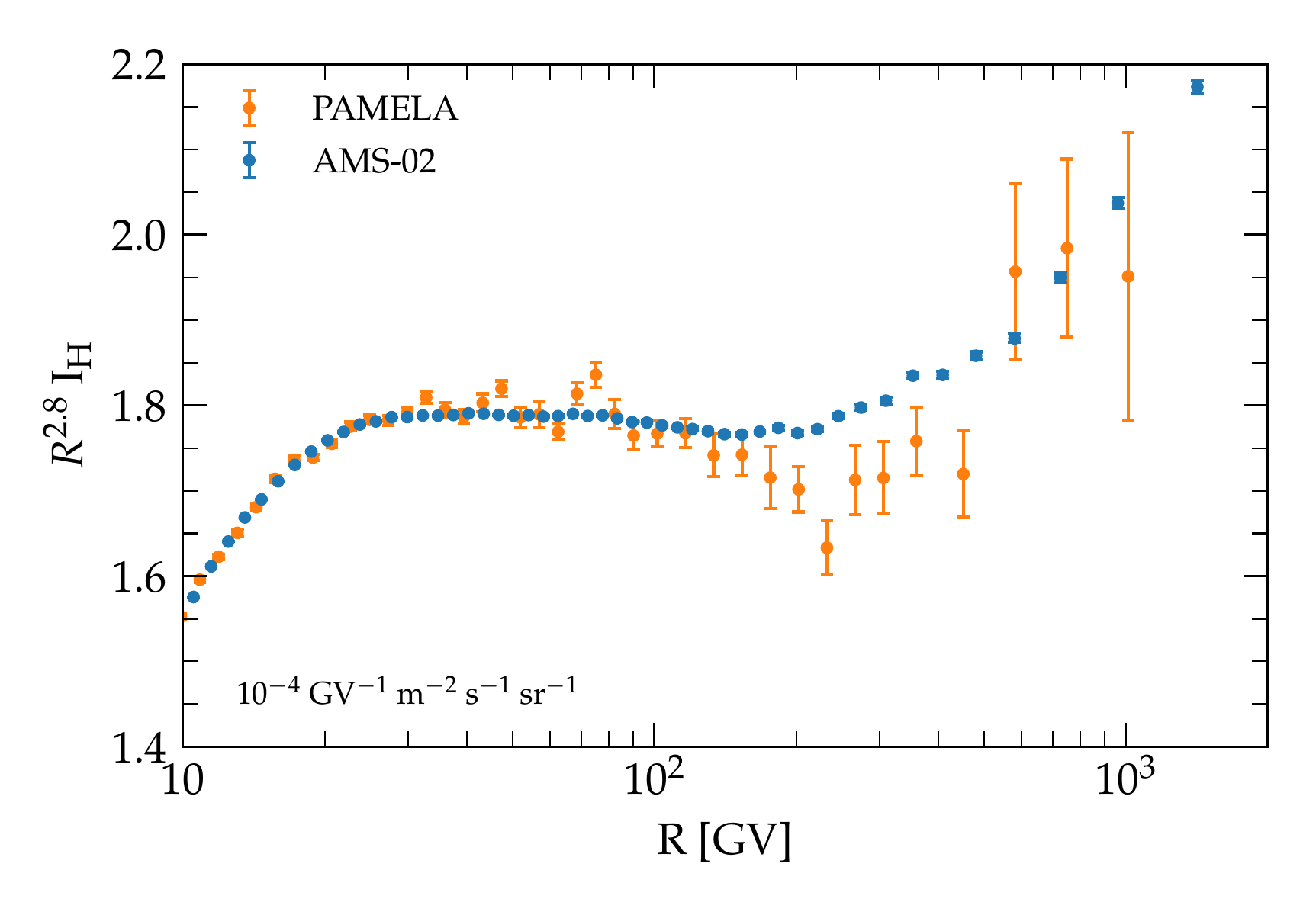}
\caption{The spectrum of protons as function of rigidity, measured by the AMS-02, and PAMELA experiments~\cite{AMS02results,PAMELA.2011.proton}.}
\label{fig:protonshe}
\end{figure}

Furthermore, by leveraging the observed slope of the proton spectrum, around 2.8 in the energy range 20-200 GeV (see figure~\ref{fig:protonshe}), we can estimate the injection slope $\gamma$ by recovering from \S\ref{sec:protons} that $\gamma + \delta \approx 2.8$, guiding us to $\gamma \approx 2.4$. 
Consequently, we arrived to the conclusion that the energy spectra of CRs at their sources must be relatively \emph{soft}, challenging the predictions based on the original version of DSA\footnote{Again, for a more comprehensive treatment of this topic, we refer readers to D.~Caprioli's lecture notes in this volume.}.

Note that equation~\eqref{eq:grammage} shows that by measuring the grammage, we can only determine the combination $H/D$, and therefore we cannot determine the normalization of the diffusion coefficient without the knowledge of the halo size.

To access the diffusion coefficient, we can make certain assumptions about the size of the Galactic halo. 
In the 1960s, it was proposed that observing radio emission from our Galaxy could provide insights into the region where CRs reside~\cite{Ginzburg1961ptps}. This is because radio emission is generated through synchrotron radiation emitted by high-energy electrons. The morphology of the radio emission indicates that it extends well beyond the Galactic disc, where magnetic fields allow particle diffusion to occur.

The region where CRs are predominantly found has a scale of approximately kiloparsecs, significantly larger than the size of the Galactic disc. By utilizing the B/C ratio to estimate $H/D$ and making assumptions about $H$, we can then determine the diffusion coefficient, $D$, and subsequently calculate the average time that CRs spend within the Galaxy, given by $H^2/D$. 

At around 10 GeV, B/C amounts roughly to $\sim 0.3$, which implies\footnote{We have employed equation~\eqref{eq:grammage} in conjunction with the relationship $2h \rho_{\rm d} = \mu_{\rm d}$.} a confinement time of approximately $\sim$100~Myr for $H \sim 3$~kpc, and it decreases as the energy increases.

To delve deeper into this topic, we require a reliable method for measuring this time. In the upcoming section, we will explore how secondary unstable isotopes, which decay with a timescale comparable to the confinement time, can serve as a cosmic clock, enabling us to estimate the Galactic residence time.

Combining this information with the grammage, which yields $H/D$, will allow us to obtain the average value of the diffusion coefficient on Galactic scales.

As we will discuss in 
\S\ref{sec:implications}, these advancements will pave the way for a theoretical comprehension of the microphysics behind the scattering and diffusion of cosmic particles by the fluctuations present in the interstellar turbulent magnetic fields.

The scrupulous reader may have noticed in figure~\ref{fig:protonshe} that the canonical slope, approximately $\sim 2.8$, does not extend up to very high energies. Instead, the CR spectrum exhibits a notable change of slope, commonly referred to as a \emph{break}, at an energy around $200$~GeV, where it hardens to about $\sim 2.6$.
At these energies, however, the equilibrium spectrum must be determined by the ratio $Q(p)/D(p)$, where both quantities are assumed to follow a pure power-law behavior. 

Prior to questioning well-established theoretical models to seek the elusive physical mechanism responsible for this bizarreness, it is crucial to ascertain whether the break should be attributed to the numerator or the denominator of equation~\eqref{eq:finalprimary}. 
In other words, we need to identify whether the break occurs at the injection stage or during the transport of CRs in the Galaxy.

Equation~\eqref{eq:bchene} quickly unravels this conundrum! 

Specifically, being dependent only on grammage, the secondary-over-primary ratio is the key parameter to examine. 
If the same change of slope is also present in the B/C ratio, then the conclusion must be that it is due to propagation, indicating a break in the diffusion coefficient. On the other hand, if no such change of slope is observed in the B/C ratio, then we must look for the solution at the injection stage.

\begin{figure}
\centering
\includegraphics[width=0.6\textwidth]{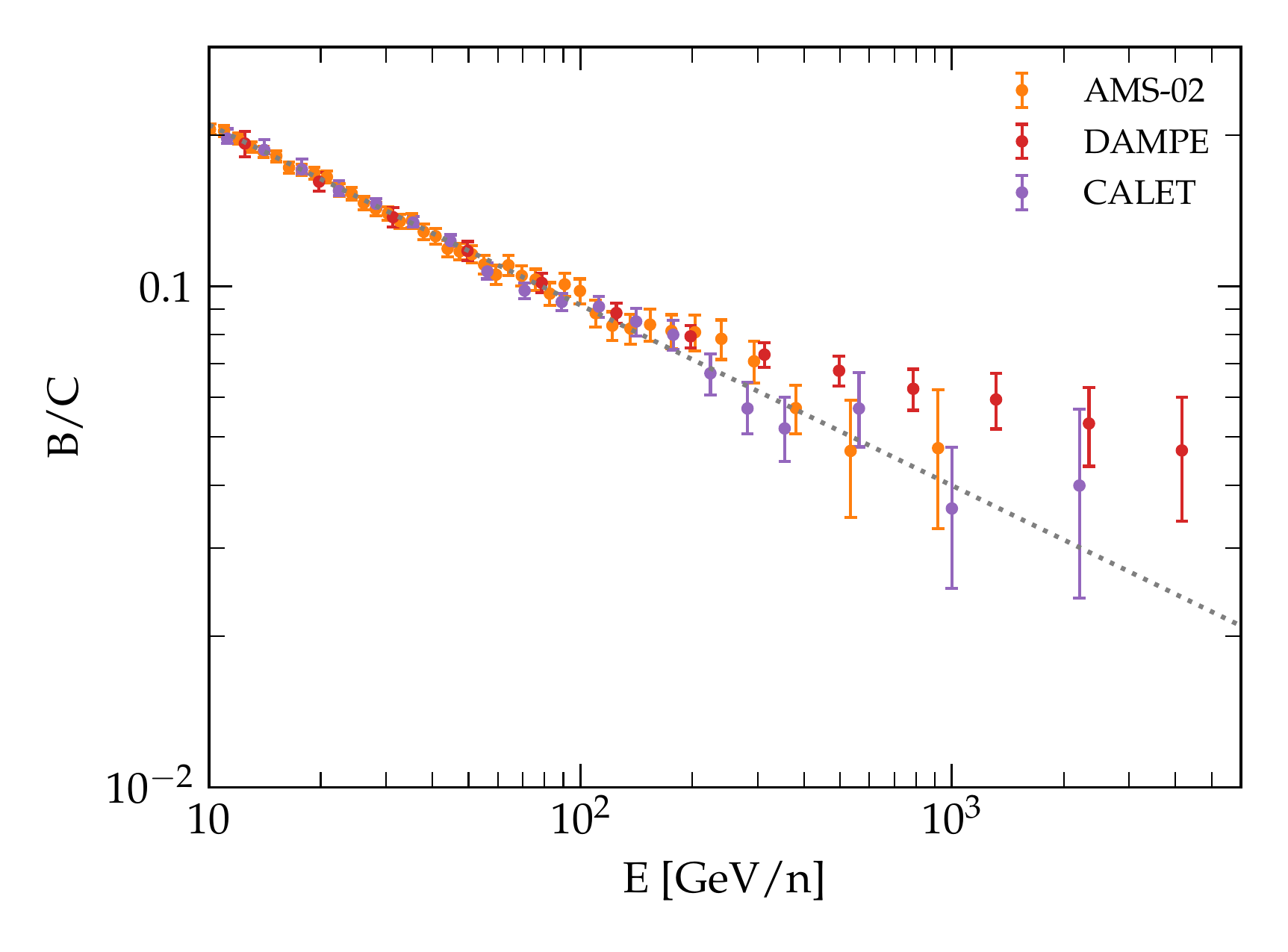}
\caption{Boron-to-carbon ratio as function of kinetic energy per nucleon measured by the AMS-02~\cite{AMS02results}, CALET~\cite{CALET.2022.BC} and DAMPE~\cite{DAMPE.2022.BC} experiments.}
\label{fig:bchighen}
\end{figure}

As shown in figure~\ref{fig:bchighen}, where we extend B/C data over the multi-TeV range thanks to measurements by DAMPE and CALET, the situation indeed aligns with the former scenario, and thereby all the current explanations of this feature are given in terms of some alteration in the galactic transport. 
These revisions in the transport of CRs might be associated with a spatial dependence of the diffusion coefficient, as proposed in ~\cite{Tomassetti2012apj}, or due to the transition from selfgenerated turbulence to preexisting turbulence, see, e.g.,~\cite{Evoli2018prl}.

At such, the recently reported departures from an otherwise boring scale-free power-law behavior in the galactic CR spectra are of paramount importance, as they offer valuable insights into the fundamental mechanisms governing the propagation of CRs in magnetized environments.

\section{Unstable nuclei: the case of beryllium}
\label{sec:unstable}

The average time that CRs spend in the Galaxy before escaping can be determined by studying the suppression of the flux of unstable nuclei due to radioactive decay. By comparing the fluxes of two isotopes of the same chemical element—one stable and the other unstable—it is possible to measure this suppression and estimate $\tau_{\rm esc}$.

Beryllium is an element that is very rare in ordinary matter, and the majority of beryllium nuclei in CRs are secondary particles formed through the fragmentation of heavier nuclei. 
It has two stable isotopes, $^9$Be and $^7$Be (if fully ionized), as well as one $\beta^-$ unstable isotope, $^{10}$Be, with a half-life of $\tau_{1/2} = (1.386 \pm 0.016)$ Myr~\cite{Chmeleff2010nimb}. 
The decay of $^{10}$Be mainly produces $^{10}$B, thus altering the abundance of this stable element.

The transport equation for $^{10}$Be can be treated as that of a secondary species, assuming (for simplicity) only one parent nucleus. 
Additionally, a term describing decay in the Galaxy reference frame is included:
\begin{equation}
-\frac{\partial}{\partial z} \left[ D_{\rm Be} \frac{\partial f_{\rm Be}}{\partial z} \right] =
- \frac{f_{\rm Be}}{\tau_{\rm f, Be}} 
- \frac{f_{\rm Be}}{\gamma \tau_{\rm d, Be}} 
+ \frac{f_{\rm C}}{\tau_{\rm f, C \rightarrow Be}} 
\end{equation}
Here, $\tau_{\rm d} = \tau_{1/2} / \ln(2)$ represents the rest-frame lifetime of $^{10}$Be, and $\gamma$ accounts for time dilation effects.
At sufficiently high energies, $^{10}$Be decays on a timescale longer than $\tau_{\rm esc}$, effectively behaving as a stable isotope.

It is worth emphasizing that this is the first case we have discussed where the source or loss term in the transport equation does not exhibit a $\delta$-function shape in $z$. This distinction arises due to the inclusion of the decay term, which introduces a new complexity that necessitates a dedicated approach for solving the corresponding transport equation.

Outside the disk $z \neq 0$, the transport equation becomes:
\begin{equation}
-\frac{\partial}{\partial z} \left[ D_{\rm Be} \frac{\partial f_{\rm Be}(z)}{\partial z} \right] + \frac{f_{\rm Be}(z)}{\gamma\tau_{\rm d, Be}} = 0
\end{equation}

Unlike the case of stable elements, the diffusive flux is not conserved. 
To find a solution, we assume the form:
\begin{equation}
f(z) = A {\rm e}^{-\alpha z} + B {\rm e}^{\alpha z}
\end{equation}
which implies $\alpha^{-1} = \sqrt{D \gamma\tau_{\rm d}}$, where we have assumed that the diffusion coefficient is spatially constant.

By imposing the appropriate boundary conditions, we obtain (introducing $y \equiv {\rm e}^{\alpha H}$):
\begin{equation}
\frac{f_{\rm Be}(z)}{f_{\rm Be,0}} =  -\frac{y^2}{1 - y^2} {\rm e}^{-\alpha z} + \frac{1}{1- y^2} {\rm e}^{\alpha z} 
\label{eq:bespatial}
\end{equation}

The value of the distribution function at $z=0$ can be obtained by integrating above and below the disk:
\begin{equation}
\left. -2 D_{\rm Be} \frac{\partial f_{\rm Be}(T)}{\partial z} \right|_{0^+} =
- 2 h n_{\rm d} c \sigma_{\rm Be} f_{0, \rm Be}
+ 2 h n_{\rm d} c \sigma_{\rm C \rightarrow Be} f_{0, \rm C} 
\end{equation}

To obtain the flux, we utilize equation~\eqref{eq:bespatial}, yielding:
\begin{equation}
\left. \frac{\partial f_{\rm Be}(T)}{\partial z} \right|_{0^+} = \alpha \frac{1+y^2}{1-y^2} \, f_{\rm Be,0}
\end{equation}

Combining these equations, we arrive at:
\begin{equation}
f_{\rm Be,0}(p) \left[  \frac{\sigma_{\rm Be}}{m_{\rm p}} - \frac{1}{c m_{\rm p} h n_{\rm d}} \sqrt{\frac{D_{\rm Be}(p)}{\gamma \tau_{\rm d, Be}}} \frac{1+y^2}{1-y^2} \right] = 
\frac{\sigma_{\rm C\rightarrow Be}}{m_{\rm p}} f_{\rm C,0}(p)
\end{equation}

Alternatively, one can write it in terms of grammages
\begin{equation}
\frac{f_{\rm Be,0}}{f_{\rm C,0}}(p) = 
\frac{1}{\rchi_{\rm cr, C \rightarrow Be}}  \left[  \frac{1}{\rchi_{\rm cr, Be}} + \frac{1}{\rchi_{\rm Be}^\prime(p)} \right]^{-1}  
\end{equation}
where $\rchi^\prime$ represents the grammage modified by the decay term and can be expressed as:
\begin{equation}
\rchi^\prime_{\rm Be}(p) = \rchi_{\rm Be}(p) \frac{1}{\alpha H} \frac{y^2 - 1}{y^2 + 1} 
\end{equation}

We observe that at high energies, when the proper decay time becomes much larger than the confinement time ($\gamma \tau_{\rm d} \gg \frac{H^2}{D}$), we have $\alpha H \rightarrow 0$ and expanding in Taylor series:
\begin{equation} 
\rchi_{\rm Be}^\prime(p) 
\oset{\gamma\tau_{\rm d} \gg \tau_{\rm esc}}{\longrightarrow} 
\rchi_{\rm Be}(p)
\end{equation}

\begin{figure}
\centering
\includegraphics[width=0.6\textwidth]{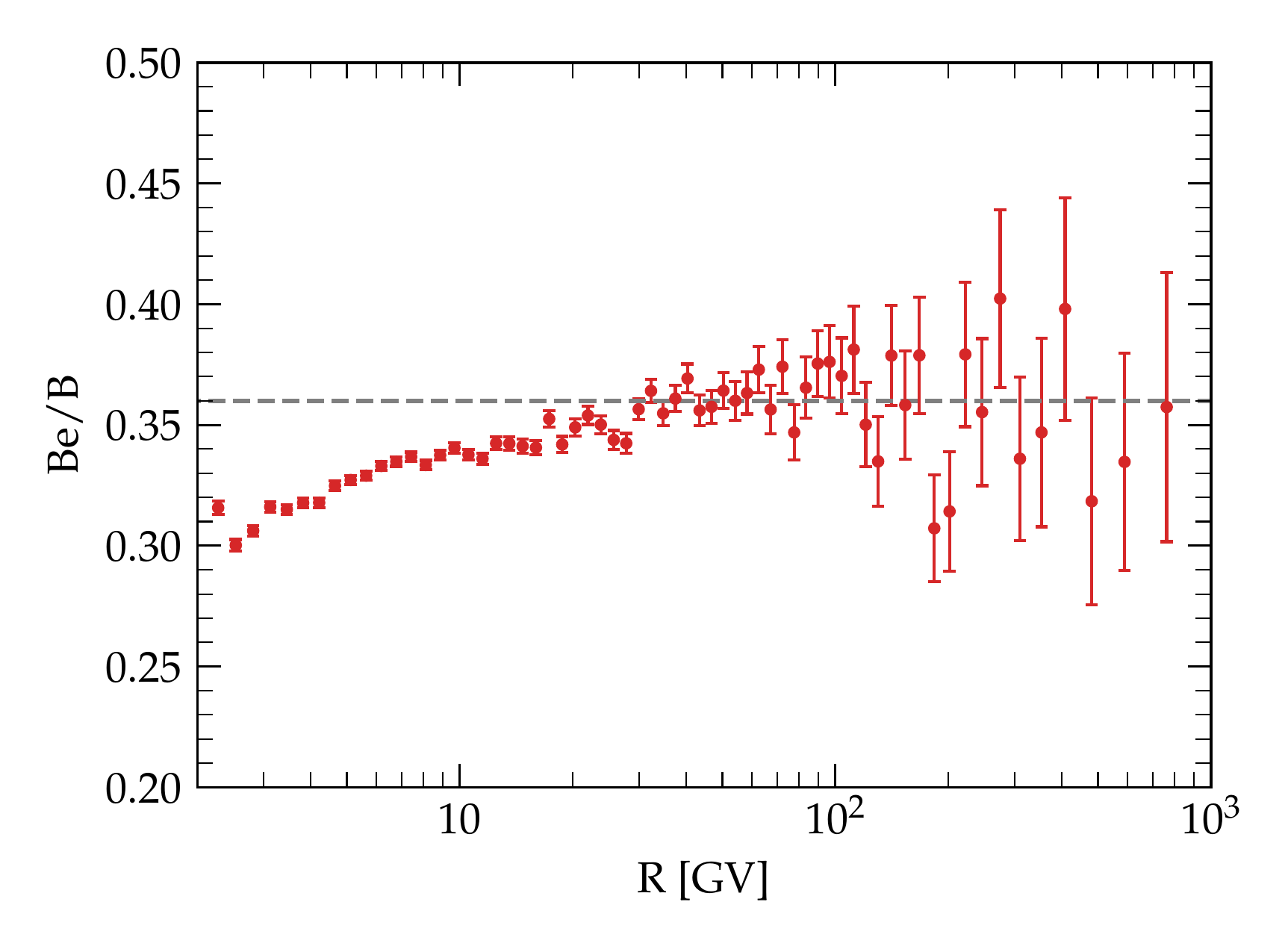}
\caption{Ratio of Beryllium over Boron fluxes as measured by AMS-02~\cite{AMS02libeb}. The dotted line shows the case without decay for 10Be.}
\label{fig:beb}
\end{figure}

As expected $\tau_{\rm d}$ cancels out from the grammage, and we recover in this limit the solution obtained for a stable element.

In the opposite limit, $y \rightarrow \infty$, and the modified grammage $\rchi_{\rm Be}^\prime(p)$ approaches a simplified form:
\begin{equation}
\rchi_{\rm Be}^\prime(p) 
\oset{\gamma\tau_{\rm d} \ll \tau_{\rm esc}}{\longrightarrow} 
m_{\rm p} n_{\rm d} h c \sqrt{\frac{\gamma\tau_{\rm d, Be}}{D_{\rm Be}(p)}} = 
m_{\rm p} \bar n c \sqrt{\gamma \tau_{\rm d} \tau_{\rm esc}} 
\label{eq:be10be9}
\end{equation}

It is important to note that in this case, it becomes crucial to account for the additional contribution to boron production arising from the decay of beryllium. This contribution can be analytically calculated using the distribution of parent beryllium in the halo, as outlined in equation~\eqref{eq:bespatial} (a detailed derivation of this contribution is provided in~\cite{Evoli2020prd}).

In the given context, the ratio of $^{10}$Be to $^9$Be fluxes can be approximated using equation~\eqref{eq:be10be9}, which relates it to the grammage ratio:
\begin{equation}
\frac{\rm {^{10}}Be}{\rm {^{9}}Be} \simeq \frac{\rchi^\prime_{\rm Be}}{\rchi_{\rm Be}} = \sqrt{\frac{\gamma \tau_d}{\tau_{\rm esc}}}
\label{eq:chibe10be9}
\end{equation}

The preliminary results from the AMS-02 experiment indicate a value of $\frac{\rm {^{10}}Be}{\rm {^{9}}Be} \sim 0.3$ at a kinetic energy per nucleon $T \sim 10$ GeV/n. Inverting the equation, we can estimate the escape time as $\tau_{\rm esc} \simeq 200$~Myr.

Furthermore, the relationship between the scale height $H$ and the escape time $\tau_{\rm esc}$ can be quantitatively expressed as:
\begin{equation}
H \sim 7 \, \text{kpc} \left(\frac{\tau_{\rm esc}}{200 \, \text{Myr}}\right) \left(\frac{\rchi}{10 \, \text{g cm}^2}\right)^{-1}
\end{equation}
roughly confirming the scaling estimated with the synchrotron emission.

While directly measuring the CR confinement time $\tau$ by comparing the flux of $^{10}$Be to that of $^9$Be remains challenging due to the difficulty of separating in mass the two isotopes, the AMS-02 experiment provides access to the total flux of beryllium, encompassing $^7$Be, $^9$Be, and $^{10}$Be, as well as the total flux of boron.
Examining the ratio of Be/B still allows us to obtain valuable information, considering that this ratio is influenced by the decay of $^{10}$Be, affecting both the numerator and the denominator (see figure~\ref{fig:beb}).

The observed behavior of the Be/B ratio at rigidities $\lesssim 30$ GV, where the ratio changes due to the decay of $^{10}$Be and not solely based on the production cross-section ratio, suggests that the escape timescale is not significantly faster than the decay timescale at that energy.

A dedicated analysis of this process, as reported in~\cite{Evoli2020prd,Weinrich2020aab,Maurin2022aa}, indicates that $H$ is approximately 6 kpc, although the possibility of a larger value cannot be excluded.

\section{Cosmic-ray transport for the poor physicists}
\label{sec:poorphysicist}

For the fast reader, it can be useful to remember a simple recipe that enables a quick retrieval of the scaling of the propagated spectrum with respect to the particle energy.

As a general rule of thumb, the equilibrium spectrum can always be obtained using the following approach:
\begin{shaded}
\begin{equation*}
\text{Intensity} \sim {\color{red}\text{Injection Rate}} \times \frac{\color{darkgreen}\text{Relevant lifetime}}{\color{blue}\text{Relevant volume}}
\end{equation*}
\end{shaded}

In the following, we recap all the cases derived so far:
\begin{itemize}
\item Primary species equilibrium spectrum:
\begin{equation*}
f_p(p) \propto {\color{red}Q(p)} \frac{\color{darkgreen}\tau_{\rm esc}(p)}{\color{blue}H}
\end{equation*}

\item Secondary stable species equilibrium spectrum:
\begin{equation*}
f_s(p) \propto  {\color{red}f_p(p) \sigma v n_{\rm d} h_d} \frac{\color{darkgreen}\tau_{\rm esc}(p)}{\color{blue}H}
\end{equation*}

\item Secondary unstable species equilibrium spectrum:
\begin{equation*}
f_s^*(p) \propto {\color{red}f_p(p) \sigma v n_{\rm d} h_d} \frac{\color{darkgreen}\tau_{\rm d}(p)}{\color{blue}\sqrt{\tau_{\rm d}(p) D(p)}}
\end{equation*}
\end{itemize}

By combining these results, we can observe the following:
\begin{itemize}
\item Stable secondary over primary ratio:
\begin{equation*}
\frac{f_s(T)}{f_p(p)} \propto \rchi(p) \propto {\color{darkorange}\frac{H}{D(p)}}
\end{equation*}

\item Unstable secondary over stable secondary ratio:
\begin{equation*}
\frac{f_s^*(p)}{f_s(p)} \propto {\color{darkorange}\frac{\sqrt{D(p)}}{H^2}}
\end{equation*}
\end{itemize}

These expressions highlight the importance of simultaneously measuring both the secondary-over-primary ratio and the unstable-over-stable ratio. Such measurements allow us to break the degeneracy between the diffusion coefficient $D$ and the size of the CR confinement volume $H$.

\section{On the leaky-box approximation to the transport equation}
\label{sec:leakybox}

In the leaky-box approximation, which provides a highly simplified version of the diffusion model, we consider a homogeneous region between two reflecting boundaries with a finite escape probability. This model, known as the \emph
{leaky-box model}, has been widely used to derive important properties of galactic CR transport, such as grammage and escape time, and has served as the foundation for much of the literature interpreting CR data~\cite{Osborne1988sval,Seo1994apj}.

By adopting this model, we can neglect any spatial dependence and express the diffusion term as a rigidity-dependent mean leakage rate, given by:
\begin{equation}
\frac{\partial}{\partial z} ( D_\alpha \nabla f_\alpha ) \longrightarrow -\frac{f_\alpha}{\tau_{\rm esc, \alpha}}(p)
\end{equation}

Here, $\tau_{\rm esc}$ represents the usual confinement time $H^2 / D$.

It is important to note that nuclear fragmentation is actually determined not by the escape time, but rather by the destruction timescale, which is related to the average gas density within the confinement volume and can be expressed as $\tau_{\rm f, \alpha} = (\bar n c \sigma_\alpha)^{-1}$.

In this simplified approach, we assume a steady-state scenario and solve the following equation:
\begin{equation}
0 = Q_\alpha(p) 
- \frac{f_\alpha}{\tau_{\rm esc, \alpha}} 
- \left( \frac{1}{\tau_{\rm f, \alpha}} + \frac{1}{\tau_{\rm d, \alpha}} \right) f_\alpha 
+ \sum_{\alpha^\prime > \alpha} \left(  \frac{1}{\tau_{\rm f, \alpha^\prime \rightarrow \alpha}} + \frac{1}{\tau_{\rm d, \alpha^\prime}} \right) f_{\alpha^\prime} 
\label{eq:lbm}
\end{equation}

In the case of stable species where $\tau_{\rm d} \rightarrow \infty$, the solution becomes:
\begin{equation}
f_\alpha = \left(Q_\alpha + \frac{f_{\alpha^\prime}}{\tau_{\rm f, \alpha^\prime \rightarrow \alpha}} \right) \left( \frac{1}{\tau_{\rm esc}, \alpha} + \frac{1}{\tau_{\rm f, \alpha}} \right)^{-1}
\end{equation}

We can rewrite this in terms of grammage as:
\begin{equation}
f_\alpha = \left( \frac{Q_\alpha}{\bar n m_{\rm p} c} + \frac{f_{\alpha^\prime}}{\rchi_{\rm cr, \alpha^\prime \rightarrow \alpha}} \right)
\left( \frac{1}{\rchi_\alpha(p)} + \frac{1}{\rchi_{\rm cr, \alpha}} \right)^{-1}
\end{equation}

For primary nuclei like carbon, where the contribution from spallation of heavier elements is negligible, equation~\eqref{eq:lbm} simplifies to:
\begin{equation}
f_{\rm C} =   \frac{Q_{\rm C}}{\bar n m_{\rm p} c} \left( \frac{1}{\rchi_{\rm C}(p)} + \frac{1}{\rchi_{\rm cr, C}} \right)^{-1}
\end{equation}

Similarly, for secondary boron produced only in spallation reactions initiated by carbon nuclei, the equation for boron nuclei becomes:
\begin{equation}
f_{\rm B} = \frac{f_{\rm C}}{\rchi_{\rm cr, C \rightarrow B}} 
\left( \frac{1}{\rchi_{\rm B}(p)} + \frac{1}{\rchi_{\rm cr, B}} \right)^{-1}
\end{equation}

This relationship implies for the B/C ratio:
\begin{equation}
\frac{\rm B}{\rm C} \simeq \frac{1}{\rchi_{\rm cr, C\rightarrow B}} \left(\frac{1}{\rchi(p)} + \frac{1}{\rchi_{\rm cr, B}}\right)^{-1}
\end{equation}

Remarkably, this equation is identical to the one we derived for the halo model in equation~\eqref{eq:chibc}.

Now let's apply the leaky-box approach to unstable elements such as $^{10}$Be. 

Considering only secondary species, the solution becomes:
\begin{equation}
f_{\rm Be} = \frac{f_{\rm C}}{\tau^{\rm in}_{\rm C \rightarrow Be}} 
\left( \frac{1}{\tau_{\rm e, Be}} + \frac{1}{\tau_{\rm f, Be}} + \frac{1}{\gamma \tau_{\rm d, Be}} \right)^{-1}
\end{equation}

Assuming similar secondary production for both isotopes ($\sigma_{10} \simeq \sigma_9$), the ratio of $^{10}$Be to $^{9}$Be can be approximated as:
\begin{equation}
\frac{\rm {^{10}}Be}{\rm {^{9}}Be} = 
\frac{\tau_{\rm esc}^{-1} + \tau_{\rm in}^{-1}}{\tau_{\rm esc}^{-1} + \tau_{\rm f}^{-1} + \gamma\tau_{\rm d}^{-1}} 
\oset{\tau^{\rm esc} \ll \tau^{\rm f}}{\longrightarrow} \left(1 + \frac{\tau_{\rm esc}}{\gamma\tau_{\rm d}}\right)^{-1}
\end{equation}

This solution must be compared to the solution we derived for the halo model in equation~\eqref{eq:chibe10be9}.

At $\gamma \sim 10$, measurements indicate a ratio of $\frac{\rm {^{10}}Be}{\rm {^{9}}Be} \sim 0.3$, which corresponds to the following timescales in the two cases:
\begin{equation} 
\tau_{\rm esc} = 
\begin{cases}
\sim 60~\text{Myr} & \text{in the Leaky-box approximation}\\
\sim 200~\text{Myr} & \text{in the thin disc model}
\end{cases}
\end{equation}

These timescales differ by more than a factor of 3! This breakdown of the leaky-box method for radioactive isotopes, where the effective volume is always smaller than the confinement volume used to determine the grammage, should serve as a warning against using this formalism for simple estimates of the escape time of CRs or the halo height.

\section{Electrons and positrons}
\label{sec:leptons}

The fraction of leptons (electrons + positrons) in the total CR flux may be small, but their unique properties make them crucial for studying fundamental astrophysical problems, such as CR propagation and the search for sources of antimatter in the Universe.

Theoretical considerations suggest that CR electrons consist of a primary component, accelerated possibly by SuperNova Remnants (SNRs) along with nuclei, and a secondary component originating from inelastic collisions between CR nuclei (mostly protons and helium) and the ISM. However, this secondary contribution accounts for only a small fraction (less than 4\%) of the total electron flux~\cite{Moskalenko1998apj,Delahaye2010aa,Evoli2021prd}.

In the standard model of CR origin, positrons were considered to be predominantly of secondary production, with a spectrum steeper than that of both primary protons and secondary nuclei (like boron) due to radiative losses. Within this framework, the positron fraction, defined as the ratio of positron flux to the sum of electrons and positrons, was expected to decrease with increasing energy. Early measurements of the positron fraction provided preliminary and intriguing evidence for a flat or even increasing trend, contradicting the standard CR transport scenario. However, the statistical limitations of these early measurements, attributed to the very low positron flux in cosmic radiation, prevented robust conclusions. Nonetheless, this finding received strong confirmation from the PAMELA experiment, which demonstrated a growing positron fraction with energy, at least up to approximately 100 GeV~\cite{PAMELA.2009.posfraction}. The excess was also observed by Fermi-LAT, which utilized Earth's magnetic field to distinguish between electrons and positrons~\cite{FERMI.2012.posfraction}. 
Subsequently, the positron excess was further confirmed and measured with higher accuracy at even higher energies by the AMS-02 experiment aboard the International Space Station. Thanks to its extended energy range, the AMS-02 experiment reported the first experimental observation of the positron fraction reaching a maximum around $\sim$500 GeV, followed by a sharp drop at higher energies~\cite{AMS02.2013.posfraction}.

In addition, the unambiguous measurement of electron and positron spectra separately revealed that the rise in the positron fraction is due to an excess of positrons rather than a deficit of electrons~\cite{AMS02.2019.electrons}.

These discoveries have unveiled a new population of sources that are primarily responsible for the production of electron-positron pairs. Among the various possibilities, galactic pulsars emerge as the most likely candidates for these leptonic sources, representing a remarkable breakthrough in our comprehension of the acceleration mechanisms occurring in these objects~\cite{Harding1987icrc,Aharonian1995aa,Grasso2009aph,Hooper2009jcap,Delahaye2010aa,Manconi2020prd,Amato2020arxiv}.

The propagation of electrons in the Galaxy is different from that of nuclei. At high energies, radiative losses become a dominant process, as loss-time scales proportionally to $E^{-1}$. Consequently, high-energy electrons have limited lifetimes and can only propagate within a restricted distance range.

For these high-energy electrons, the primary energy loss mechanisms are synchrotron radiation resulting from interactions with interstellar magnetic fields and inverse Compton scattering (ICS) with Galactic radiation fields (such as the cosmic microwave background, infrared, optical, and UV photons).

In the Thompson approximation, neglecting Klein-Nishina corrections to the $\gamma$-e$^-$ cross-section\footnote{In the context of the Galaxy, this assumption encounters some limitations, particularly concerning the ICS of electrons with optical and UV photons. A more accurate approach to ICS reveals spectral features arising from the diminishing Klein-Nishina cross-section of electrons as their energy surpasses around $\sim 40$~GeV when interacting with UV photons~\cite{Agaronyan1985ap,vanderwalt1991mnras,Evoli2020prl}.}, the rate of energy loss can be expressed as follows (see P.D.~Serpico's lecture notes in this volume):
\begin{equation}
\left| \frac{dE}{dt} \right| = \frac{4}{3} \sigma_{\rm T} c \gamma^2 \beta^2 (\mathcal U_\gamma + \mathcal U_{\rm B}) = b_0 \left( \frac{E}{10~\rm GeV} \right)^2
\end{equation}
Here, $\sigma_{\rm T}$ represents the Thomson cross section, $\mathcal U_\gamma$ denotes the energy density in background photons, and $\mathcal U_{\rm B} = \frac{B^2}{8\pi}$ represents the magnetic field energy density.

In Galactic environments, the energy densities $\mathcal U_i$ typically range from $\mathcal O(0.1-1~\text{eV/cm}^3)$, leading to a value of $b_0$:
\begin{equation}
b_0 \sim  10^{-14} \left(\frac{\mathcal U_\gamma + \mathcal U_{\rm B}}{\rm eV/cm^3}\right) \left(\frac{E}{10~\rm GeV}\right)^2 \, \text{GeV} \, \text{s}^{-1}
\end{equation}

The energy loss time turns out to be a \emph{decreasing} function with energy:
\begin{equation}
\tau_{\rm loss} \simeq \frac{E}{-dE/dt} \sim 30 \, \text{Myr} \left(\frac{E}{10~\rm GeV}\right)^{-1}
\end{equation}

\begin{figure}[t]
\centering
\includegraphics[width=0.6\textwidth]{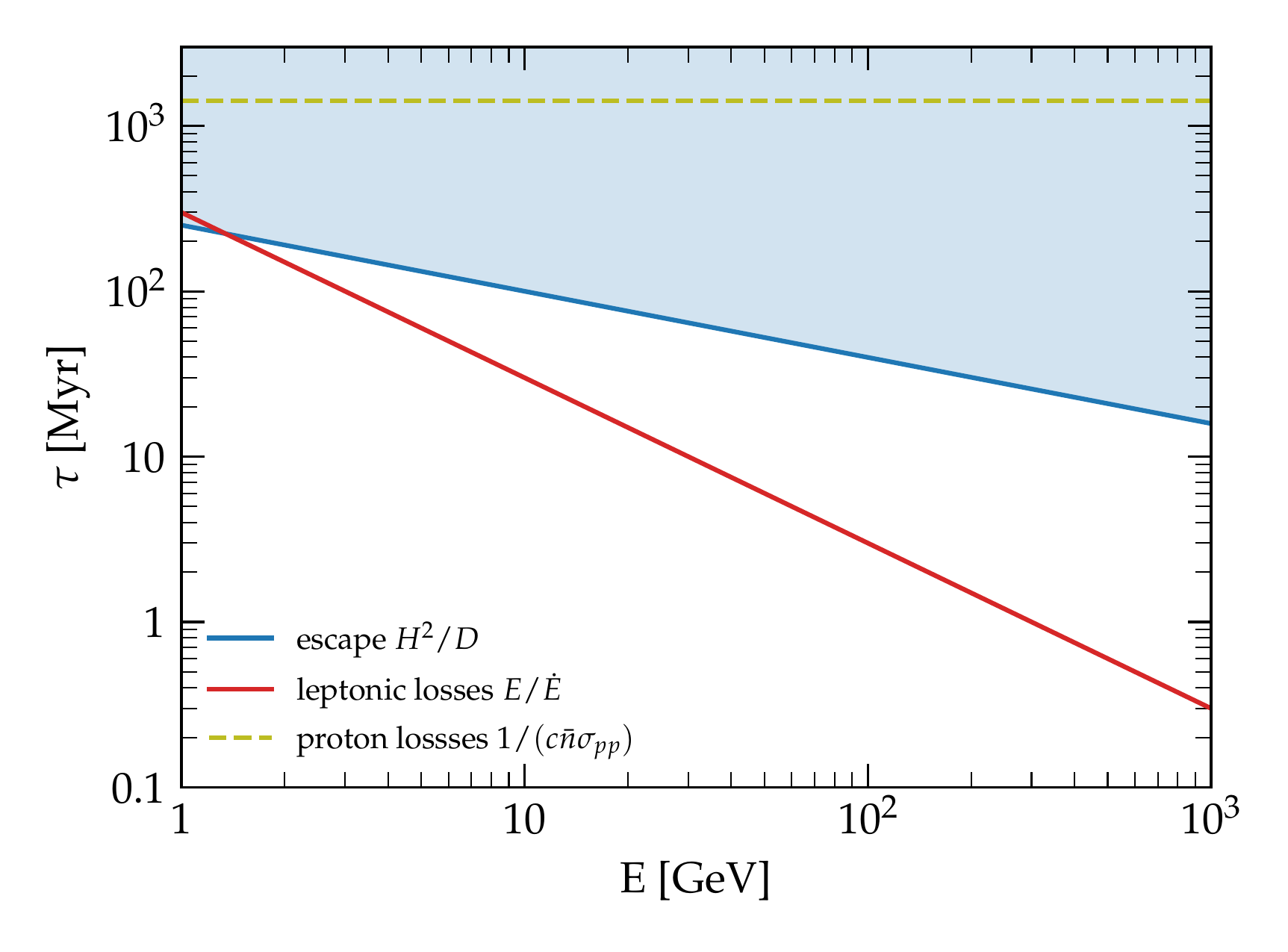}
\caption{The escape timescale derived in \S\ref{sec:protons} is confronted with energy loss timescales for protons (inelastic scattering) and leptons (synchrotron + IC) in the Milky Way.}
\label{fig:electronlosses}
\end{figure}

Figure~\ref{fig:electronlosses} provides a comparison between the energy loss timescale for electrons and the CR escape timescale as derived from nuclei. For typical values of CR transport in the ISM, the transition between the two regimes occurs at a few GeV. This implies that the Galaxy acts as an effective calorimeter for leptons, as they are expected to lose a significant portion of their energy during the typical escape time.

The transport equation for leptons is described by\footnote{For CR electrons $p \simeq E$}:
\begin{equation}
-\frac{\partial}{\partial z} \left [D \frac{\partial f_e}{\partial z} \right] 
= Q_e(E) \delta(z)
- \frac{1}{E^2} \frac{\partial}{\partial E} \left[ \dot E E^2 f_e \right] 
\end{equation}
where $\dot E$ represents the energy loss rate.

To simplify the loss term, it is convenient to approximate it as a catastrophic loss term~\footnote{Notice that assuming $\dot E \propto E^2$ and $f_e \propto E^{-\alpha}$ the relative difference between the two terms is $\sim\alpha$.}:
\begin{equation}
-\frac{\partial}{\partial z} \left [D \frac{\partial f_e(E)}{\partial z} \right] = Q_e(E) \delta(z) - \frac{f_e(E)}{\tau_{\rm loss}(E)}  
\label{eq:leptonprop}
\end{equation}
allowing the equation to be solved similarly to unstable nuclei, as the energy losses are effective throughout the propagation volume.

In the limit of negligible losses, the solution of equation~\eqref{eq:leptonprop} corresponds to that of a stable species. On the other hand, when losses dominate transport ($\tau_{\rm loss} \ll \tau_{\rm esc}$), the solution can be expressed as:
\begin{equation}f_{e,0}(E) 
= \frac{Q_{e,0}(E) \mathcal R_{\rm SN}}{2 \pi R_{\rm d}^2} \frac{\tau_{\rm loss}(E)}{\sqrt{D(E)\tau_{\rm loss}(E)}}
\propto E^{{-\gamma}-\frac{1 + \delta}{2}}
\label{eq:leptonsolution}
\end{equation}

\begin{figure}[t]
\centering
\includegraphics[width=0.6\textwidth]{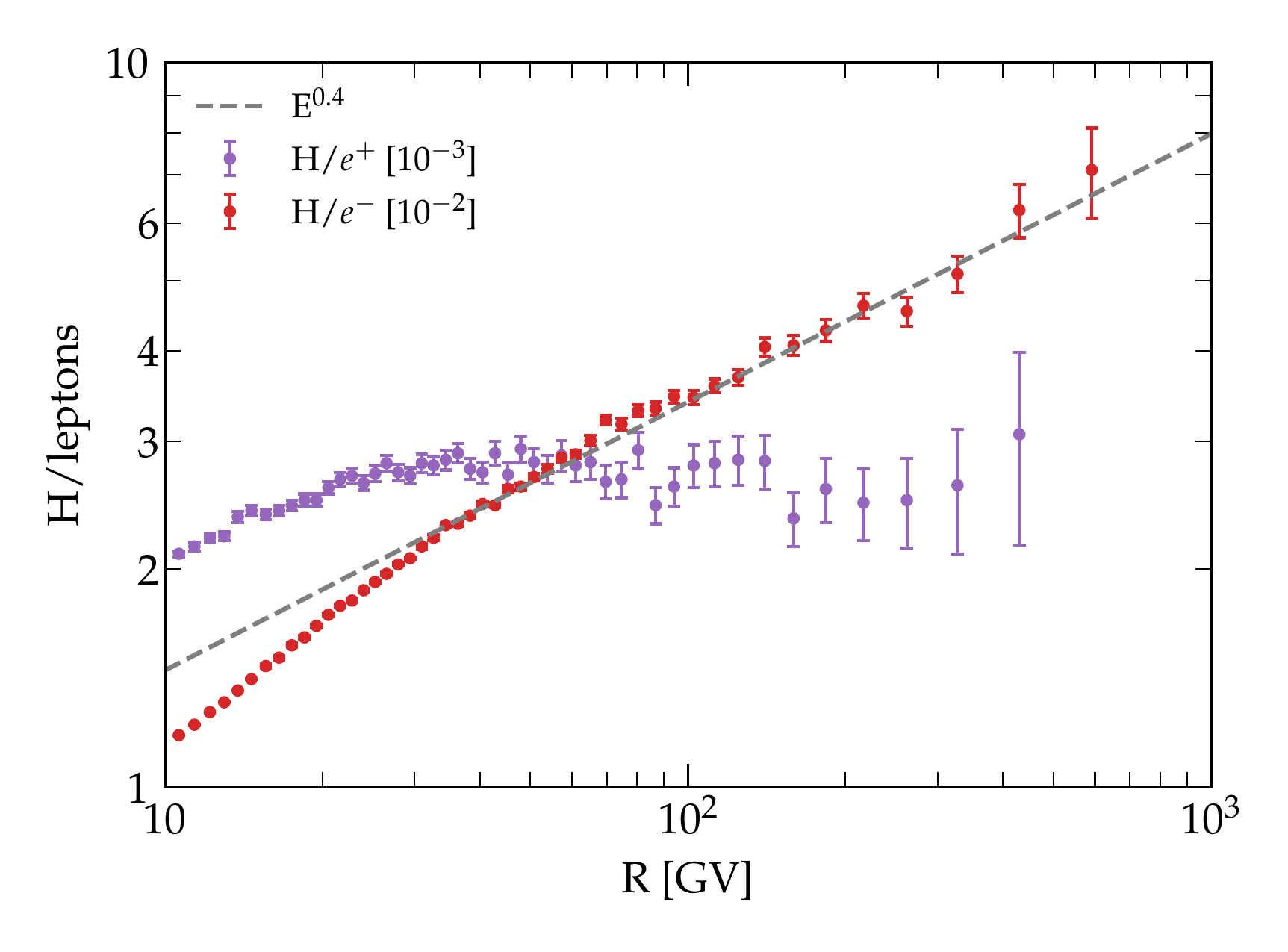}
\caption{The electron(positron)-over-proton ratio as measured by AMS-02~\cite{AMS02.2019.electrons}. The high-energy power law fit is also shown as a dashed line.}
\label{fig:protonelectron}
\end{figure}

Thereby, the transition from a diffusion-dominated regime to a losses-dominated regime results in a \emph{softening} of the electron spectrum. This transition leads to a change in slope of
\begin{equation}
\Delta \alpha = (-\gamma-\delta) - (-\gamma-\frac{1 + \delta}{2}) = \frac{1 -\delta}{2} \simeq 0.3
\end{equation}

In Galactic CRs, the spectral break is not easily discernible due to the strong influence of solar modulation in the energy range where the break is expected to occur. 
Moreover, above the energy range where solar modulation plays a significant role, the most prominent feature in the electron spectrum is the spectral steepening at energies $E \lesssim$ TeV. The spectral break in the electron spectrum is well-described by a broken power-law with a change of slope of approximately $\sim$1~\cite{HESS.2008.leptons,DAMPE.2017.leptons,CALET.2018.leptons}, which is too large to be attributed to the \emph{cooling} break resulting from the transition between the diffusion-dominated and losses-dominated regimes\footnote{However, refer to~\cite{Cowsik1979apj,Lipari2017prd} for instances where unconventional approaches are explored, continuing to challenge the foundational principles outlined in these lecture notes.}. 
This further strengthens the evidence that electron transport is predominantly governed by energy losses throughout the entire energy range.

When comparing the proton and electron spectra, see figure~\ref{fig:protonelectron}, the difference in normalization is likely caused by the different injection mechanisms into the acceleration process for electrons and protons, as suggested by previous studies~\cite{Morlino2021mnras}.

On the other hand, we may be tempted to attribute the steeper slope of the electron spectrum solely to their energy losses.
Between 50 and 500 GeV, the ratio of proton flux to electron flux is well described by a power law with a slope of $\sim$0.4.

According to the prediction of the diffusion-losses model, this ratio can be expressed as:
\begin{equation}
\frac{f_p}{f_e} \propto \frac{E^{-\gamma_p-\delta}}{E^{{-\gamma_e}-\frac{1 + \delta}{2}}} \propto E^{-(\gamma_p-\gamma_e)} E^{\frac{1-\delta}{2}}
\end{equation}
where we differentiate between the injection spectra of protons ($\gamma_p$) and electrons ($\gamma_e$).

By comparing the proton-over-electron ratio with observational data, as shown in the figure, we find $\Delta \gamma = \gamma_e-\gamma_p \simeq 0.1$. This suggests that the \emph{injection} spectrum of electrons is relatively steep. More accurate analyses, accounting for realistic energy losses in the Galaxy, have found even larger values of $\Delta \gamma \simeq 0.3$~\cite{Evoli2021prd}. Resolving this issue is challenging since the most probable explanation, namely that the electron spectrum is steepened by losses in the downstream region of a SNR shock, requires extreme conditions in the late stages of SNR evolution~\cite{Cristofari2021aa}. Consequently, the origin of the steeper electron spectrum remains an open question.

Additionally, the efficiency of energy losses introduces a characteristic propagation scale, denoted as $l \simeq \sqrt{D(E) \tau_{\rm loss}}$, which serves as an effective \emph{horizon} defining the maximum distance from which an electron source of energy E can contribute to the flux observed at Earth.

Quantitatively, this scale is approximately given by
\begin{equation}
\frac{l}{H} \simeq \sqrt{\frac{\tau_{\rm loss}}{\tau_{\rm esc}}} \simeq 0.6 \, \left(\frac{E}{10\,\text{GeV}}\right)^{-\frac{1+\delta}{2}}
\end{equation}

Due to the existence of this horizon, only sources within a distance where the propagation time is shorter than the loss time at that energy can significantly contribute to the observed flux. Assuming a uniform distribution of sources within the Galactic disk, the estimated number of sources exploding in a loss timescale $\tau_{\rm loss}$ and lying within a distance $l$ from Earth is given by 
\begin{equation}\label{eq:nleptons}
N(E) \simeq \frac{\mathcal R \tau_{\rm loss} l^2(E)}{R_{\rm d}^2} \simeq 50 \left(\frac{E}{\rm TeV}\right)^{-2 + \delta}
\end{equation}

This simple estimation highlights the rapid decrease in the number of contributing sources with increasing energy, making the high-energy spectrum highly sensitive to the precise distribution of sources in our galactic vicinity.

What we learn from this is that while we routinely assume a homogeneous distribution of sources in the galactic disk, in reality, CR sources exhibit discrete spatial and temporal characteristics. As the number of sources approaches unity ($N \sim 1$), the discrete nature of the sources becomes increasingly relevant. This is in contrast to nuclei, where the spectrum is weakly dependent on the exact distribution of sources in space and time, as protons and nuclei diffuse over kiloparsec scales before escaping the CR halo, effectively averaging over the distribution of sources on these scales.

As a consequence of equation~\eqref{eq:nleptons}, it is plausible that the lepton flux in the multi-TeV energy range may receive a significant contribution from a local source. Consequently, the detection of such a source becomes an attainable goal for ongoing experiments like DAMPE and CALET, which aim to explore this energy range in the near future~\cite{Evoli2021prdb}.

\begin{figure}[t]
\centering
\includegraphics[width=0.6\textwidth]{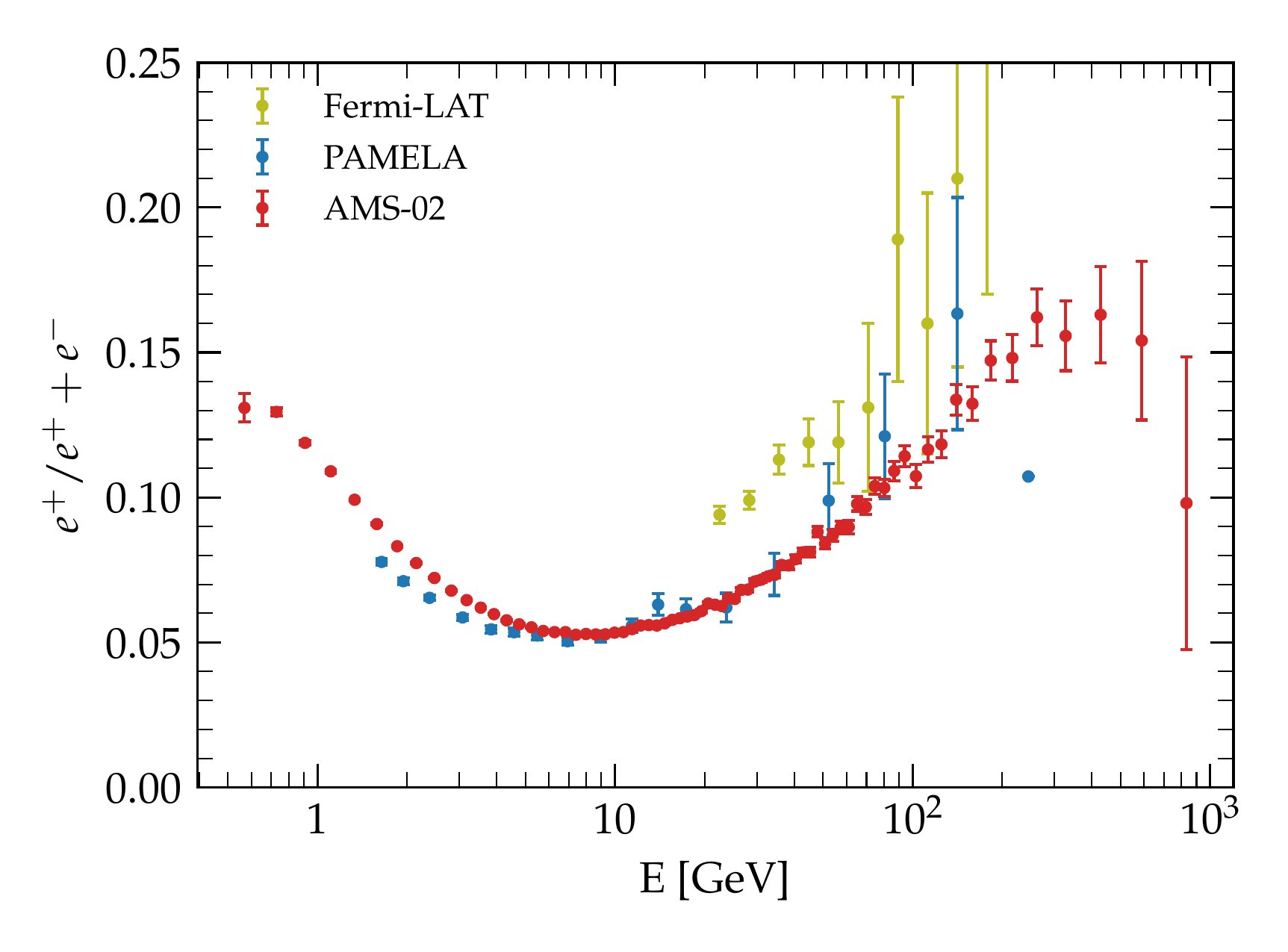}
\caption{The positron fraction as a function of electron or positron energy as measured by PAMELA, FERMI, and AMS-02~\cite{PAMELA.2009.posfraction,FERMI.2012.posfraction,AMS02.2013.posfraction}.}
\label{fig:positronfraction}
\end{figure}

It is interesting to apply this model to compute secondary electrons and positrons.

Secondary positrons are primarily produced through nuclear reactions between protons in the cosmic radiation and protons in the target gas, resulting in the production of charged pions ($\pi^\pm$) and other mesons, with positrons being one of the final products of the decay chain.

Typically, the energy of secondary positrons is a fraction $\xi \sim \mathcal{O}(5\%)$ of the parent proton energy $E_p$:
\begin{equation}
E_{e^+} \simeq \xi E_p 
\end{equation} 

The rate of positron ($e^+$) production in the ISM can be expressed as:
\begin{equation}
q_{e^+}(E) dE = n_p(E_p) dE_p \sigma_{\rm pp} c 2 h_d n_{\rm d} \delta(z) 
\label{eq:positronproduction}
\end{equation}

Applying the solution of equation~\eqref{eq:leptonprop}, when losses are unimportant, we obtain:
\begin{equation}
f_{e^+}(E) = 
n_p\!\left(\frac{E}{\xi} \right) \frac{2 c \sigma_{\rm pp} n_d h_d}{\xi}  \frac{H}{D(E)} 
\end{equation}

Whereas, in the limit where losses dominate, equation~\eqref{eq:leptonsolution}, we have:
\begin{equation}
f_{e^+}(E) = 
n_p\!\left(\frac{E}{\xi} \right) \frac{2 c \sigma_{\rm pp} n_d h_d}{\xi}  \frac{\tau_{\rm loss}(E)}{\sqrt{\tau_{\rm loss}(E) D(E)}} 
\end{equation}

It is worth noting that the proton spectrum is always evaluated at an energy $1/\xi$ larger than the positron energy.

In both cases, one obtains:
\begin{equation} 
\frac{f_{e^+}}{f_{e^-}}(E) = \frac{q_{p,0}(E/\xi)}{q_{e,0}(E)} \frac{1}{\xi} \frac{\rchi(E / \xi)}{\hat\rchi} \sim E^{-\gamma_p+\gamma_e-\delta}
\label{eq:positronfractiontheory}
\end{equation}

\begin{figure}[t]
\centering
\includegraphics[width=0.6\textwidth]{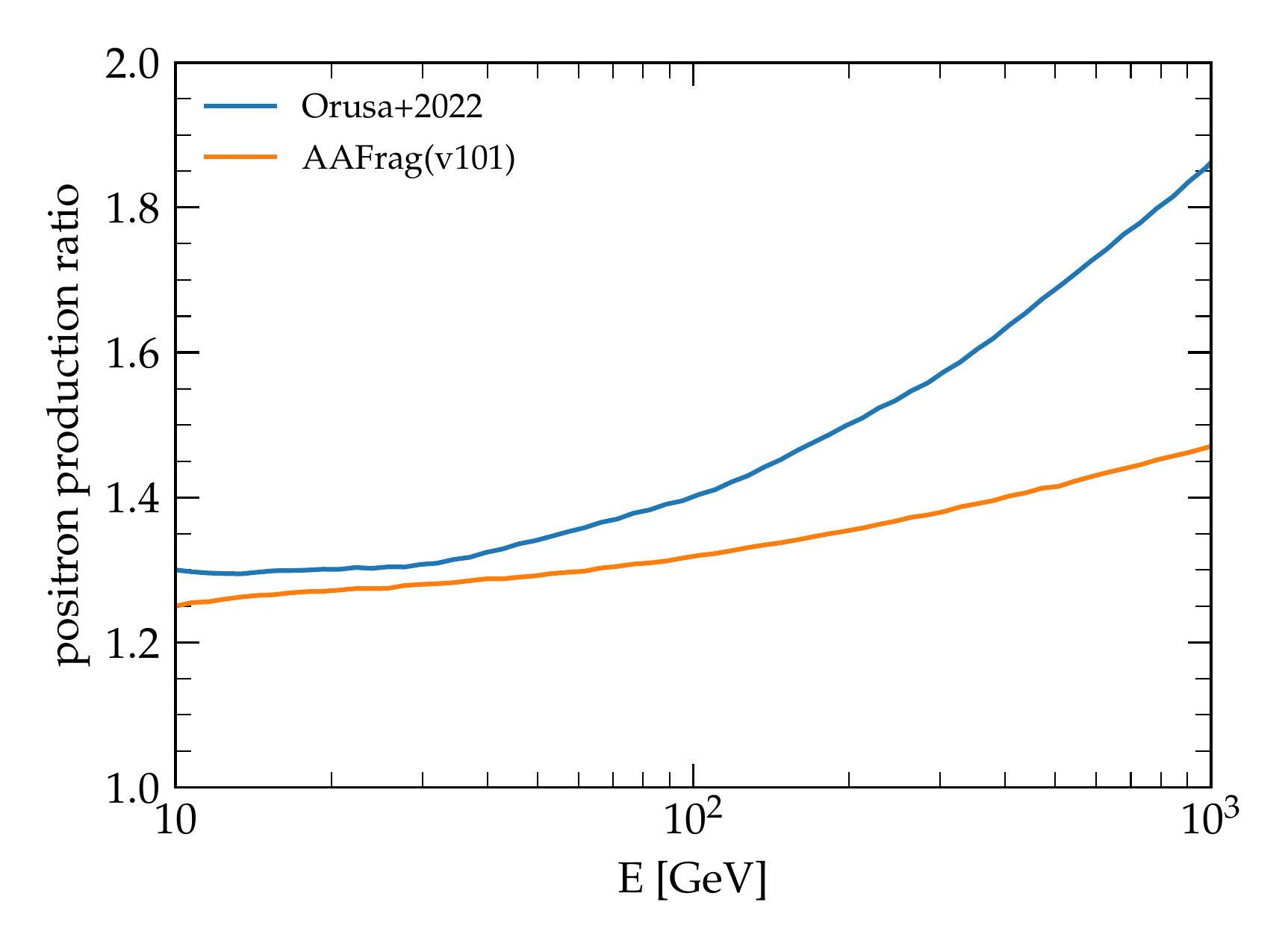}
\caption{The positron production rate computed in two recent models~\cite{Kachelriess2019cpc,Orusa2022prd} normalized to the production rate in the constant-\emph{inelasticity} approach outlined in the text.}
\label{fig:positronsourceratio}
\end{figure}

This is different from the case of B/C, where carbon was the parent of boron,  as here electrons are not parents of secondary electrons.
Notwithstanding, assuming $\gamma_p \simeq \gamma_e$, the positron fraction is a decreasing function with energy, approximately following $E^{-\delta}$.

In figure~\ref{fig:positronfraction}, we present the positron fraction as a function of energy, which markedly deviates from the expectation for pure secondary production above $\sim$10 GeV.

Following equation~\eqref{eq:positronfractiontheory}, for the positron fraction to increase with energy, it would require $\gamma_e > \gamma_p + \delta$, which is highly unlikely!

Another possibility we may consider to explain the anomaly in the positron fraction is a significant modification of the cross-sections involved in secondary production processes.
Recent efforts have been made to re-evaluate these cross-sections by fitting data from collider experiments or by utilizing hadronic interaction models~\cite{Orusa2022prd,Kachelriess2019cpc}. The production rates obtained using these approaches can be compared with the rates predicted by equation~\eqref{eq:positronproduction}, as shown in figure~\ref{fig:positronsourceratio}. 
The comparison reveals that there are no deviations from our initial naive approach at a level that would account for the observed excess.

As such, we are left with no other option than to postulate the existence of a new population of positron sources in the Universe!

\section{Immediate implications of cosmic ray observations}
\label{sec:implications}

\subsection{Efficiency of particle acceleration in Galactic sources}

In the previous sections, we have discussed how the abundances of certain elements such as boron, lithium and beryllium in CRs provide us with valuable estimates of the time $\tau_{\rm esc}$ that CRs spend in the Galaxy before escaping.
Now, we delve deeper into the implications of these observations, specifically focusing on the energetic budget required by galactic sources to sustain the CR population.

Having in mind acceleration mechanisms similar to DSA, to describe the injection spectrum of protons, we assume a power-law form in momentum that accounts for both relativistic and non-relativistic particles:
\begin{equation}
N(p) = N_0 \left(\frac{p}{m c}\right)^{-\gamma} \, , 
\end{equation}
where $\gamma \gtrsim 4$. The normalization of $N(p)$ is determined by the condition that the integrated energy in particles matches the energy released in CRs by a single event:
\begin{equation}
4 \pi \int_0^\infty dp \, p^2 N(p) T(p) = E_{\rm CR}
\end{equation}

Solving for $N_0$, we find
\begin{equation}
N_0 = \frac{E_{\rm CR}}{4 \pi c (m c)^4 I(\gamma)},
\end{equation}
where $I(\gamma) = \int_0^\infty dx \, x^{2-\gamma} \left[ \sqrt{x^2+1} - 1 \right]$.

Note that due to spectral index values larger than 4, the total energy budget is determined by protons with energies of $\sim$GeV, and we can ignore the existence of minimum and maximum momentum.

Assuming high energies where ionization losses can be neglected and solar modulation has no significant effect, the proton spectrum contributed by identical sources occurring at a rate $\mathcal R$ can be expressed as:
\begin{equation}
f_{\rm p}(p) = \frac{E_{\rm CR} \mathcal R}{8 \pi^2 R_{\rm d}^2 c (m c)^4 I(\gamma)} \left( \frac{p}{mc}\right)^{-\gamma} \frac{H}{D(p)},
\end{equation}

Using the definition of intensity, given in appendix~\ref{app:intensity}, and considering that in the relativistic limit $E \simeq p c$, we obtain:
\begin{equation}
I_{\rm p}(E) = \frac{E_{\rm CR} \mathcal R c}{8 \pi^2 R_{\rm d}^2 (m c^2)^2 I(\gamma)} \left( \frac{E}{mc^2}\right)^{2-\gamma} \frac{H}{D(E)}
\end{equation}
which gives for $E = 10$~GeV:
\begin{equation}
E^2 I_{\rm p}(E) \simeq 2 \times 10^3 \left(\frac{E_{\rm CR} \mathcal R}{10^{40} \, \text{erg} \, \text{s}^{-1}}\right) \, \text{GeV} \, \text{m}^{-2} \, \text{s}^{-1} \, \text{sr}^{-1}
\end{equation}

\begin{figure}[t]
\centering
\includegraphics[width=0.6\textwidth]{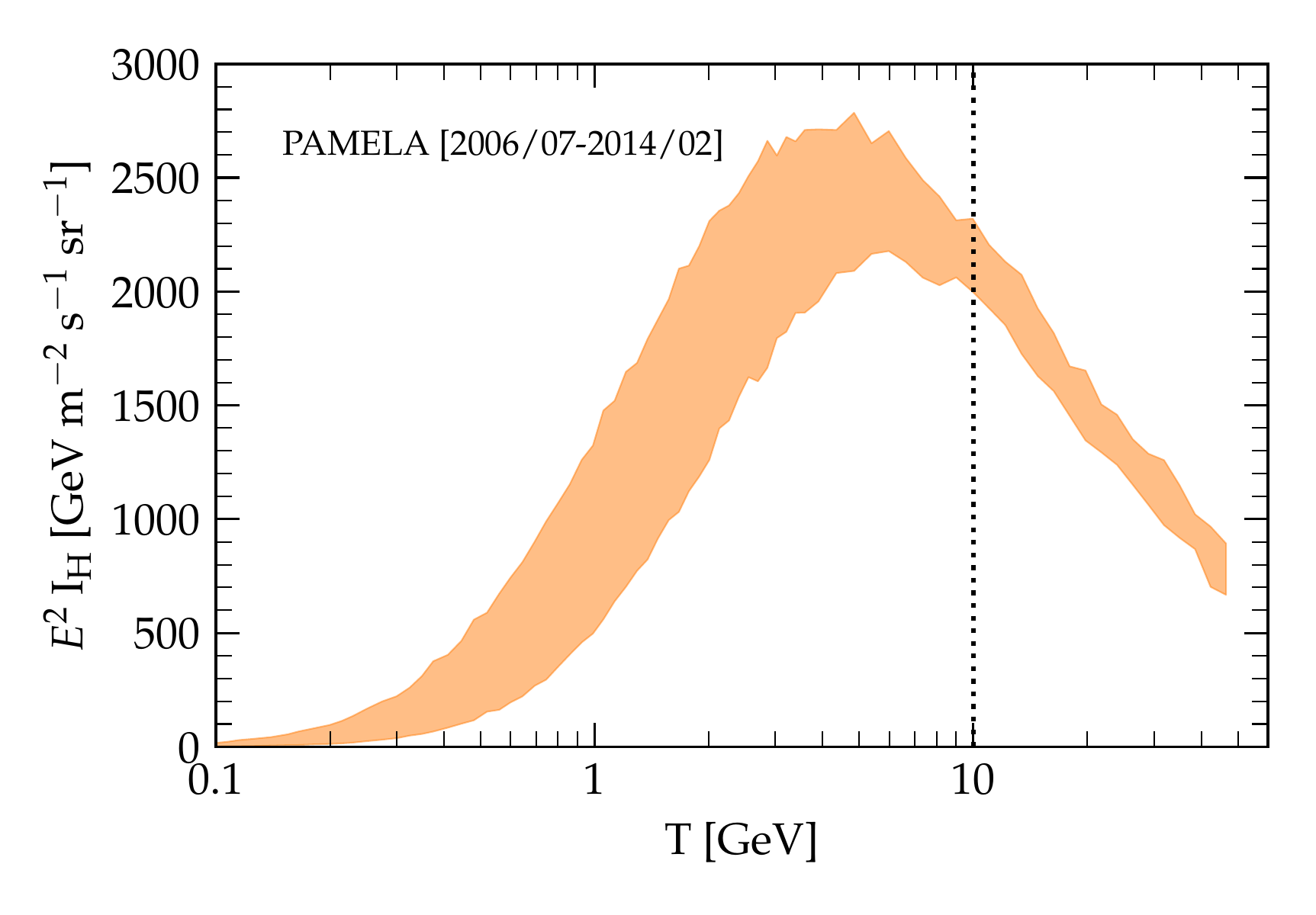}
\caption{The proton intensity measured by the PAMELA experiment over a large fraction of the Solar activity cycle ~\cite{PAMELA.2011.proton}.}
\label{fig:pamelaprotons}
\end{figure}

By comparing this equation with the proton flux measured by the PAMELA experiment, as shown in figure~\ref{fig:pamelaprotons}, we find that in order to maintain a steady-state, the power that Galactic sources inject into the Galaxy in the form of CR protons needs to be approximately $\mathcal L \simeq E_{\rm CR} \mathcal R \simeq 10^{40} \text{erg} \, \text{s}^{-1}$ for a time not smaller than $\tau_{\rm esc}$.

Assuming that CRs are produced by supernova explosions with a rate of about 2 per century and with a typical mechanical energy release of $10^{51}$ erg per explosion, the luminosity amounts to $6 \times 10^{41}$ erg/s, significantly exceeding the required one. Hence, it is necessary to invoke an efficiency of a few percent in the conversion between supernova kinetic energy and CR energy to make this hypothesis viable.

Detailed calculations provide a more accurate estimate of the total acceleration efficiency, typically ranging from 5\% to 10\% (including nuclei), for the majority of supernova remnants. This efficiency is well described in recent models of diffusive shock acceleration~\cite{Morlino2017hsn}, upraising the hypothesis that CRs acquire their energy from Galactic stellar explosions to the rank of a \emph{paradigm}.

Over the years, alternative sources of energy, such as those in the Galactic Center region, star clusters, or OB associations\footnote{OB associations consist largely of very young, massive stars (about 10 to 50 solar masses) of spectral types O and B.}, have been proposed to explain galactic CRs. 
Interestingly, the star clusters scenario for the origin of galactic CRs have recently gained renewed attention based on gamma-ray observations (see S.~Gabici's lecture notes in this volume).


\subsection{Constraints on the microphysics of galactic transport}

In the previous sections, we have discussed how the remarkable longevity of CRs, which greatly exceed the time it would take them to propagate freely at the speed of light, suggests that CRs undergo a random walk, continually scattered as they traverse their path from the sources to Earth.
Furthermore, the absence of a strong anisotropy, $\mathcal{O}(1)$, toward the galactic center for CRs with energies below $10^{15}$ eV implies that multiple scatterings wash out the anisotropy that would be expected if there were no scattering~\cite{Hillas2005jpg}. 

The scattering cannot be attributed to ISM nuclei, as the mean free path for Coulomb collisions of relativistic nucleons in the dilute ISM, given by $\lambda \simeq \frac{1}{n_{\rm H} \sigma_{\rm T}} \sim 10^{24}$ cm, is far too long.

Hence, the most likely scattering mechanism is the interaction of CRs with plasma waves, which are fluctuating electromagnetic fields in the ISM. The presence of a random component of the Galactic magnetic field is inferred from fluctuations in rotation measures, combined with estimates of thermal electron density and synchrotron depolarization. Recent analyses have reported a strength of this component at around $1-3 \mu$G~\cite{Ferriere2023epj}.

Understanding the production of these plasma waves and their influence on the dynamics of CR particles is a crucial topic in the theoretical investigation of CR physics.

Typically, the energy spectrum of the turbulent field is assumed to follow a power-law distribution with an outer scale of $L \sim 10$ pc, where energy is injected. This energy then cascades down to smaller scales until it dissipates at the dissipation scale.

For CR nuclei in the GeV-TeV energy range and charge $Z$, the Larmor radius in this magnetic field is given by:
\begin{equation}
r_{\rm L} \simeq 10^{-5} \, \text{pc} \left( \frac{E / Z}{10 \, \text{GeV}} \right) \left( \frac{B}{\mu\text{G}}\right)^{-1}  
\end{equation}

This length scale is much smaller than the random field injection scale but larger than the dissipation scale. Consequently, a CR nucleus is expected to encounter numerous magnetic scattering centers before reaching Earth.

This transport mechanism is understood in terms of resonant wave-particle interaction, where CRs primarily scatter off waves with wavelengths comparable to their gyroradius. 

Under these conditions, a particle interacts with a wave remaining in phase with the wave over many cycles. 
This scattering leads to an effective diffusion in pitch angle which, in turn, regulates their diffusion in real space.

The resonant condition can be expressed as
\begin{equation}
k_{\rm res} = \frac{1}{\mu r_{\rm L}}
\end{equation}
where $k_{\rm res}$ represents the wavelength of the \emph{resonant} magnetic perturbation and $\mu$ is the cosine of the particle's pitch angle.

Within quasilinear theory the spatial diffusion coefficient results from the pitch-angle-cosine $\mu$ average of the inverse of the pitch-angle diffusion coefficient $D_{\mu\mu}$ (see appendix~\ref{app:diffusioncoefficient}):
\begin{equation}
D \simeq \frac{(1-\mu^2) v^2}{D_{\mu\mu}}
\label{eq:dandbasta}
\end{equation}

The pitch-angle diffusion rate must depend upon the distribution of wave energy, and in weakly turbulent magnetic fields, we obtain:
\begin{equation}
D_{\mu\mu} \simeq \pi \Omega (1-\mu^2) k_{\rm res} \mathcal F(k_{\rm res})
\label{eq:dmumusimeq}
\end{equation}
where $\Omega = c / r_{\rm L}$ is the gyrofrequency, and $\mathcal F$ is the power in modes of wavenumber $k_{\rm res}$ normalized such that its integral gives the fraction of energy density in the turbulent components with respect to the regular field:
\begin{equation}
\int_{1/L}^\infty \! dk \, \mathcal F(k) = \frac{\langle \delta B^2 \rangle}{B_0^2}
\end{equation}

Intuitively, if we associate the angle by which the fieldlines are bent $\delta B/B_0$ with the scattering angle $\delta \theta$, and assume the particles encounter uncorrelated waves at frequency $\Omega$, then the angular diffusion coefficient is $\langle (\delta \theta)^2 \rangle / \delta t \sim \Omega (\delta B/B_0)^2$ which is essentially equation~\eqref{eq:dmumusimeq}. 

Combined with equation~\eqref{eq:dandbasta} the diffusion coefficient reads
\begin{equation}
D \simeq \frac{1}{3} r_{\rm L} c \frac{1}{k_{\rm res} \mathcal F(k_{\rm res})}
\label{eq:dzz}
\end{equation} 

We recall now from equation~\eqref{eq:grammage} that the best fit to secondary-over-primary CRs in the Galaxy indicates a diffusion coefficient of $D/H \sim 2$ for particles around 10 GeV, where $D$ is measured in units of $10^{28}$ cm$^2$ s$^{-1}$ and $H$ represents the scale height in kpc. 
Combining this information with the determination of the halo size obtained from unstable species, $H \sim 5$ kpc, we obtain a \emph{measurement} of the normalization of the diffusion coefficient to be approximately $D \sim 10^{29}$ cm$^2$ s$^{-1}$.


By inverting equation~\eqref{eq:dzz}, we can finally determine the required level of turbulence at the scale of 10 GeV particles in order to replenish the observed amount of secondary particles:
\begin{equation}
k_{\rm res} \mathcal F(k_{\rm res}) \simeq \text{few} \times 10^{-6}
\end{equation}

In summary, even such a small perturbation at a scale corresponding to the size of the solar system, $\sim$A.U., is enough to significantly extend the transport time of CRs in the Galaxy from thousands of years to millions of years!

\section{The Green function formalism}
\label{sec:green}
In the previous sections, our calculations were based on the assumption that we can approximate the distribution of CR sources in the Galaxy as continuous in both space and time within a volume that represents the overall structure of the Galaxy.

However, it is important to note that Galactic CRs are actually accelerated at discrete sources, and we do not have detailed knowledge of the individual positions and ages of these sources. Instead, we can only infer statistical properties about them.

While the continuous source approximation is valid when the density of sources is sufficiently high in both space and time, allowing us to describe them as a continuum, it fails when the transport distances and times become comparable to or shorter than the typical separations and ages of the sources. This is particularly relevant for low-energy nuclei (affected by ionization and Coulomb energy losses) and high-energy electrons (subject to inverse Compton and synchrotron losses), for which the discrete nature of the sources needs to be taken into account.

This raises the question of how we can deduce the statistical distribution of the predicted CR flux from the statistical properties of these discrete sources, as well as whether the fluctuations in CR density are significant. 

The objective of this section is to introduce the fundamental principles of this theory by developing a method for constructing the statistical quantities of interest, such as the expectation value of the flux at Earth, through the ensemble average over similar galaxies with randomly distributed galactic sources~\cite{Lee1979apj,Ptuskin2006asr,Evoli2021prdb}.

Specifically, the statistical characteristics of the quantities under investigation will be expressed in terms of a Green's function that describes the propagation of particles from a discrete monoenergetic galactic source, and of a formal expression for the probability density of CR sources at a given point in space.

In order to have a model that is relatively simple and amendable to calculations, we consider the diffusion equation for protons with a spatially constant diffusion coefficient, denoted as $D$, and neglect any energy losses. The diffusion equation takes the form:
\begin{equation}
\frac{\partial n(\vb r, E, t)}{\partial t} - D(E) \nabla^2 n(\vb r, E, t) = Q(\vb r, E, t)
\end{equation}

To find a suitable Green's function, denoted as $\mathcal{G}$, we seek a solution that satisfies the following equation:
\begin{equation}
\frac{\partial}{\partial t} \mathcal G(\vb r, t \leftarrow \vb r_\star, t_\star) 
- D \nabla^2 \mathcal G(\vb r, t \leftarrow \vb r_\star, t_\star) 
= \delta^{(3)}(\vb r - \vb r_\star) \delta(t - t_\star)
\end{equation}
where $\vb r_\star$ and $t_\star$ represent the position and time of injection of a CR particle. The boundary conditions for $\mathcal{G}$ are the same as those assumed for $n(\vb r, t)$.

In this formalism, the Green's function $\mathcal{G}(\vb r, t \leftarrow \vb r_\star, t_\star)$ represents the probability for a CR injected at position $\vb r_\star$ and time $t_\star$ to propagate through the Galaxy and reach an observer located at $\vb r$ at time $t$.

The formal solution of the diffusion equation can then be expressed as the convolution of the Green's function with the source term over the Galactic volume and over the past Galactic history as
\begin{equation}
n(\vb r, t) = \int_0^{\infty} \! dt_\star \! \int_{V} \!  d^3 \vb r_\star \, \mathcal G(\vb r, t \leftarrow \vb r_\star, t_\star) \, Q(\vb r_\star, t_\star)
\label{eq:greenconv}
\end{equation}

In particular, we can replace the continuous and smooth source term, $Q(\vb r, t)$, with an ensemble of $N$ sources with distances $\{ \vb r_i \}$ and ages $\{ t_i \}$: 
\begin{equation}
Q(\vb r_\star, t_\star) = \sum_{i=0}^N Q_0 \delta^{(3)}(\vb r_\star - \vb r_i) \delta(t_\star - t_i)
\label{eq:qsources}
\end{equation}

Notice that, for simplicity, we assume in this expression a common, time-independent, spectrum for all sources, whereas the Green function formalism would easily allow us to take into account an arbitrary temporal evolution of the injection of CRs and the variety of source properties.

Equation~\eqref{eq:qsources} implies that the total flux can be expressed as the sum of the individual fluxes from each source as:
\begin{equation}
n(\vb r, t) = Q_0 \sum_{i=0}^N \mathcal G(\vb r, t \leftarrow \vb r_i, t_i)
\end{equation}

However, since we lack knowledge about the actual positions and ages of the sources, we must resort to a Monte Carlo approach. In this approach, a large ensemble of $\mathcal{M}$ galaxies is simulated. In each realization $\alpha$ of the ensemble, the properties of each source are randomly sampled. It is important to note that, by adopting this approach, the CR density $n_\alpha$ behaves as a stochastic variable, allowing us to analyze its statistical properties.

To validate the Monte Carlo approach, we aim to demonstrate that the conventional CR model can be recovered in the \emph{mean field limit}, where we average the flux over the ensemble of all possible realizations.

To achieve this, we consider an ensemble of similar galaxies in which the $N$ sources are distributed according to the probability $\tilde P(\mathbf{r}_1, t_1; \mathbf{r}_2, t_2; \dots; \mathbf{r}_n, t_n)$, which needs to be specified. Here, $\tilde P$ represents the probability of a specific realization, characterized by ${ \mathbf{r}_i, t_i }$, to be obtained.
Hence, the normalization condition for this probability distribution is given by:
\begin{equation}
1 = \int d\mathbf{r}_1 dt_1 d\mathbf{r}_2, dt_2 \dots d\mathbf{r}_n, dt_n \tilde P(\mathbf{r}_1, t_1; \mathbf{r}_2, t_2; \dots; \mathbf{r}_n, t_n)
\end{equation}

Consequently, the ensemble-averaged density of protons at the Sun's position can be expressed as the convolution of the density, computed given a specific source configuration, and the probability of that configuration to be realized:
\begin{multline}
n = \langle n_\alpha(\vb r_\odot, t_\odot) \rangle_\alpha = \\
= \int \! d\vb r_1 dt_1 \dots d\vb r_n dt_n \tilde P(\vb r_1, t_1; \dots; \vb r_n, t_n) n_\alpha(\vb r_\odot, t_\odot; \vb r_1, t_1; \dots; \vb r_n, t_n)
\end{multline}

Assuming that sources are not correlated with each other, we can write this as:
\begin{equation}
n = Q_0 N \int_0^{\infty} \! dt_\star \! \int_{V} \! d^3 \vb r_\star \, P(\vb r_\star, t_\star) \, {\mathcal G(\vb r_\odot, t_\odot \leftarrow \vb r_\star, t_\star)} 
\label{eq:greenn}
\end{equation}

Notice that the problem of modeling galactic CRs naturally splits in two independent components: one part describes the CR transport and its geometry, embedded in the Green function $\mathcal G$, while a second part accounts of the injection, $Q_0$, and the distribution, $P$, of sources.

The probability distribution for a constant rate of injection and spatially homogeneous sources is given by:
\begin{equation}
P = \frac{1}{N}\frac{\mathcal{R}}{\pi R_d^2}
\end{equation}

We remind that in the case of the pure diffusive problem, the Green's function would simply be (see appendix~\ref{sec:appgreen}):
\begin{equation}
\mathcal{G}(\vb r_\odot, t_\odot \leftarrow \vb r, t) = \frac{1}{(4\pi D \tau)^{3/2}} 
\exp \left[ -\frac{r^2}{4 D \tau} \right] 
\sum_{n=-\infty}^{+\infty} (-1)^n \exp \left[ -\frac{(2nH)^2}{4 D \tau} \right]
\label{eq:gdiffdisk}
\end{equation}
where $\tau = t_\odot - t$ is the time elapsed between the injection and the observation of a CR.

As we are describing a homogeneous model, in equation~\eqref{eq:gdiffdisk}, we choose the observer position at the center of the disc, i.e., $\vb r_\odot = (0,0,0)$, without loss of generality. Additionally, we consider the disc to be infinitely thin in the $z$ direction, so the sources lie in the plane $z = 0$.

We now have all the necessary components to compute the mean Galactic density. Evaluating the integral expression in equation~\eqref{eq:greenn}, we have:
\begin{equation}
n = \frac{Q_0 \mathcal R}{\pi R_d^2}
\int_0^{\infty} \!\! \frac{d\tau}{(4\pi D \tau)^{3/2}} \int_{0}^{R_d} 2 \pi r_\star^2 d r_\star
\exp \left[ -\frac{r_\star^2}{4 D \tau} \right]  
\sum_{n=-\infty}^{+\infty} (-1)^n \exp \left[ -\frac{(2nH)^2}{4 D \tau} \right]
\end{equation}

Performing the integrals, first with respect to $\tau$ and then with respect to $r_\star$, we obtain:
\begin{equation}
n = \frac{Q_0\mathcal R}{2\pi D R_d} \sum_{n=-\infty}^{+\infty} (-1)^n 
\left[ \sqrt{1 + \left(\frac{2 n H}{R_d}\right)^2} - \sqrt{\left(\frac{2 n H}{R_d}\right)^2} \, \right]
\end{equation}

If $H \ll R_{\rm d}$, it is easy to show that the sum over $n$ tends to $H/R_d$. Therefore, we have:
\begin{equation}
n =\frac{Q_0 \mathcal R}{2\pi R_{\rm d}^2}\frac{H}{D}
\end{equation}

This result demonstrates that we recover the solution of the diffusion equation in steady state in equation~\eqref{eq:protonsimplesolution}.
The mean value of the density from randomly distributed point sources is equal to the steady-state flux obtained with a continuous source distribution. This is also true in more general cases such as thick disk, finite radius, with wind and spallation.

This situation is reminiscent of the one encountered in thermodynamics, where the values of observable (macroscopic) quantities are calculated without knowledge of the details of microscopic states.

An important advantage of introducing this approach is the ability to calculate all the higher moments of the CR observables, providing us with a comprehensive understanding of the statistical properties associated with CR densities.

In particular, we are interested in the \emph{spread} of the density around its average value. 

To compute the variance of the CR density, we can utilize a property of uncorrelated random variables with a common mean and variance. Let us consider the sum, $\psi$, of $N$ uncorrelated variables, $\phi_i$, drawn from the same probability density with a mean, $\mu$, and variance, $\sigma^2$.

As a random variable, $\psi$ has a mean given simply by:
\begin{equation}
\langle \psi \rangle = N \mu
\label{eq:meanpsi}
\end{equation}

Taking advantage of the uncorrelatedness of $\phi_i$, the variance of $\psi$ can be expressed as $\sigma^2(\psi) = N \sigma^2$, which can be written using the mean value in the equation above:
\begin{equation}
\sigma^2(\psi) = N \langle \phi^2 \rangle - \frac{\langle \psi \rangle^2}{N} \simeq N \langle \phi^2 \rangle
\label{eq:sigmapsi2}
\end{equation}

In this case, the last term is neglected, assuming that the number of sources, $N$, is sufficiently large.

It is important to note that equation~\eqref{eq:meanpsi} was implicitly employed in equation~\eqref{eq:greenn}, while equation~\eqref{eq:sigmapsi2} is used here to determine the variance, as follows:
\begin{equation}
\sigma^2 \left[ n(\vb r, t) \right] =
Q_0^2 N  
\int_0^{\infty} \! dt_\star \! \int_{V} \! d^3 \vb r_\star
P(\vb r_\star, t_\star) 
\mathcal G^2(\vb r, t \leftarrow \vb r_\star, t_\star)
\end{equation}

To proceed, we make use of the \emph{free} Green's function derived in appendix~\ref{sec:appgreen}. 
Extending the generalization to incorporate the halo function would necessitate additional, intricate calculations.

Therefore, the variance becomes:
\begin{equation}
\sigma^2 \left[ n(\vb r, t) \right] =
\frac{Q_0^2 \mathcal{R}}{\pi R_d^2} 
\int_0^{\infty} \! \frac{d\tau}{(4\pi D \tau)^{3}} \! \int_0^{R_{\rm d}} \! 2 \pi r_\star^2 dr_\star 
\exp \left[ -\frac{2  r_\star^2}{4 D \tau} \right] 
\end{equation}

As before, we first integrate over $\tau$, which yields:
\begin{equation}
\sigma^2 \left[ n(\vb r, t) \right] =
\frac{Q_0^2 \mathcal{R}}{8 \pi^3 R_d^2 D} 
\! \int_0^{R_{\rm d}} \! \frac{dr_\star}{r_\star^2}
\end{equation}

We immediately notice that the spatial integral \emph{diverges} unless a lower bound is imposed on the minimum distance from the closest source, denoted as $R_{\rm min}$, such that the variance, after applying this regularization, becomes:
\begin{equation}
\sigma^2 \left[ n(\vb r, t) \right] =
\frac{Q_0^2 \mathcal{R}}{8 \pi^3 R_d^2 D R_{\rm min}} 
\end{equation}

Finally, the strength of the fluctuation is given by:
\begin{equation}
\delta n (\vb r, t)  = \frac{\sqrt{\sigma^2}}{n} = \frac{1}{\sqrt{2\pi}} \frac{R_{\rm d}}{H} \frac{D^{1/2}}{\mathcal R^{1/2} R_{\rm min}^{1/2}}
\end{equation}

This result highlights the strong dependence of $\delta n$ on the cutoff radius $R_{\rm min}$ that we introduced to avoid the divergence in the integral. Moreover, if $R_{\rm min}$ is naively chosen as a fixed value, then $\delta n$ becomes an increasing function with energy proportional to $D^{1/2}$.

It must be noted that the calculation related to the standard deviation of this quantity diverges when we allow for the possibility of sources that are extremely close ($r_\star \rightarrow 0$) and young ($\tau \rightarrow 0$), which contribute to the total density with $n \rightarrow \infty$.

As the standard deviation is commonly interpreted as the typical spread of random values around the mean, a high standard deviation implies that the actual value of the flux has a disturbingly high probability of being very far from the mean value.

One might argue that the problem we considered is physically irrelevant because there are no sources with zero age and null distance to the Earth. We can impose a lower cutoff on ages and distances, based on observations for instance, to eliminate these very rare events that lead to a very high standard deviation without significantly contributing to the mean value.

However, the level of variance is highly sensitive to this cutoff, and it is not clear what value should be adopted~\cite{Mertsch2011jcap}. This issue has important implications for the theoretical limitations in the extraction of propagation parameters and for comparing theoretical models with current experimental precision~\cite{Genolini2017aa}.

In fact, it has been noted that the divergence arises from a long power law tail in the probability density function for the density from individual sources. Hence, this difficulty is not as severe as it initially seems because meaningful statistical quantities, such as confidence levels or quantiles, can still be computed in this limit. This makes the problem tractable again, and it turns out that the total flux is distributed following a stable distribution~\cite{Bernard2012aa}.

This situation, where some rare events have a very small contribution to the mean but result in a very high standard deviation, is not uncommon in physics and arises in various contexts.

\section{Conclusions}
\label{sec:conclusions}
These notes started off in \S~\ref{sec:prelude} with an interesting anecdote about the beginning of the CR field and the link to the Varenna International School of Physics. Subsequently, they rested upon the crucial observation from our solar neighborhood (see \S~\ref{sec:intro}) contrasting the isotopic composition of the local CR flux with that of the gaseous surroundings in the solar system. Notably, the relative abundance of elements like Lithium, Beryllium, and Boron - elements that are not expected to be produced except through CRs - indicates that a lot happens between the time CRs are generated at the source and the moment they are detected. This implies a complex propagation history for each CR constituent while diffusing throughout our Galaxy. Recalling CR composition, we have seen that CRs can be classified as either primary, originating from the same ISM particles accelerated to relativistic energies, or secondary, produced during propagation thanks to different mechanisms. The secondary component is insightful in confirming the diffusive propagation of CRs across the galaxy and gives us hints about processes at play, including interaction with the galactic large-scale magnetic field, spallation, and radiative decay. A crucial role is played by grammage, which represents the amount of material a particle goes through and is an indicator of the frequency of interaction with the ISM material, and is thus directly related to the CR propagation time.

Let us now recall the key points we have pointed out throughout these lecture notes. We stress again the assumptions at the basis of every CR transport model: high-energy particles are produced and accelerated in sources in the disk with a spectrum proportional to $E^{-\gamma}$, then propagate in the Galactic halo following a diffusive motion with a diffusion coefficient proportional to $E^{\delta}$, and finally escape freely at the boundaries. Proceeding along these lines we first derived in \S~\ref{sec:protons} the spectrum of CR protons by solving the diffusion equation, resting upon the assumptions of both stationarity and a spatially constant diffusion coefficient. It is important to emphasize that the observed spectrum, proportional to $E^{-\gamma-\delta}$, is always steeper than the injected one. This observation alone does not provide sufficient insights to disentangle between acceleration effects, the $\gamma$ index, and the propagation ones, the $\delta$ index. In \S~\ref{sec:nuclei} we introduced spallation, a key mechanism consisting of fragmentation of ISM nuclei that occurs while CRs propagate and becomes relevant when a critical value for grammage is reached. An important measured quantity, as discussed in \S~\ref{sec:secondaryoverprimary}, is the secondary-over-primary ratio in both regimes of strong and weak spallation, as it helps us to estimate a value for the diffusion coefficient, provided that an assumption about the halo size has to be made. In \S~\ref{sec:unstable} we showed how secondary unstable isotopes, such as Berillium, can serve us as a cosmic clock to estimate the CR residence time and thus allow us to assess both the value of the diffusion coefficient and that of the halo size. After a quick recap of the general form of CR spectrum (see \S~\ref{sec:poorphysicist}) and the limitations of the leaky box approximation to the transport equation (see \S~\ref{sec:leakybox}), finally in \S~\ref{sec:leptons} we investigated the leptonic component, i.e. electrons and positrons. The observed positron fraction poses numerous theoretical challenges up to postulating the existence of a new population of sources of antimatter in the Universe. Last but not least, in \S~\ref{sec:implications}, we delved into some considerations about the micro-physics behind the transport process and we concluded with \S~\ref{sec:green} showing the basics of the Green's function formalism.

The outlined phenomenological model captures the fundamental observations and highlights the key physical processes CRs experience while diffusing within our Galaxy. Notwithstanding, this framework extends beyond our Galaxy's confines, with applications to a wide array of astrophysical environments.

\appendix
    
\section{The cosmic ray intensity}
\label{app:intensity}

In this appendix, we introduce the various definitions used to describe CR density. 

The number of particles in a volume element $d^3 \vb{r}$ about $\vb r$ and in the momentum interval $d^3 \vb p$ about $\vb p$ is given by
\begin{equation}
d\rho = \phi(\vb r, \vb p, t) \, d^3 \vb r \, d^3 \vb p
\end{equation}
with $\phi$ the distribution function.

Expanding $d^3 \vb p$ in spherical coordinates gives:
\begin{equation}
d\rho  = \phi(\vb r, p, t) \, d^3 \vb r \, p^2 dp \, d\Omega
\end{equation}

Typically we are not able to measure $\phi$ but only averages over momentum space, thereby we conveniently introduce the \emph{phase-space distribution function} as:
\begin{equation}
f(\vb r, p, t) = \frac{1}{4\pi} \int_\Omega \phi(\vb r, p, t) d\Omega
\end{equation}

Correspondingly, the number of particles in $d^3 \vb r$ and in $dp$ (independent of direction of $\vb p$) is:
\begin{equation}
dn = \int_\Omega \! d\Omega \, F(\vb r, p, t) \, d^3 \vb r \, p^2 dp = 4 \pi p^2 f(\vb r, p, t) \, d^3 \vb r \, dp
\end{equation}
 
Consequently, when the spectral index of the CR phase distribution in momentum is represented by $\gamma$, the corresponding slope, when expressed in terms of energy, becomes $\gamma - 2$:
\begin{equation}
f \propto p^{-\gamma} \, \rightarrow \, n \propto E^{2-\gamma}
\end{equation}

Experimental measurements typically present their results in terms of the CR differential intensity, denoted as $I$. This quantity represents the number of particles collected per unit energy, unit area, unit time, and unit steradians.

Furthermore, it is convenient to describe the spectra of the different CR components as a function of the kinetic energy per nucleon, denoted as $T$. This choice is advantageous because the kinetic energy per nucleon is approximately conserved when a nucleus fragments due to interactions with the interstellar gas.

In terms of the phase-space distribution, the intensity of a specific nuclear species $\alpha$, can be expressed as 
\begin{equation}
I_\alpha (T) dT = c p^2 f_\alpha(p) \beta (p) dp 
\end{equation}
which can be further simplified as 
\begin{equation}
I_\alpha(T) = A p^2 f_\alpha(p)
\label{eq:f2I}
\end{equation}
where $A$ is the nuclear mass.

\section{Calculation of diffusion coefficient in quasilinear-theory}
\label{app:diffusioncoefficient}

In this appendix, we explore the interaction between charged particles and an astrophysical plasma to derive the spatial diffusion coefficient using the quasilinear theory (QLT). 
QLT allows to directly compute this coefficient and other transport parameters based on the previous knowledge of the turbulent spectra.
The quasilinear approximation can be seen as a first-order perturbation theory and here we follow standard derivations as in~\cite{Blandford1987pr,Shalchi2009book,Blasi2013aar}.

Recent developments on this subject are given in the P.~Blasi and A.~Marcowith lecture notes in this volume.

First, we consider the equations of motion for a particle traveling within an ordered magnetic field aligned with the $\hat{\vb z}$ axis, denoted as $\vb B_0 = B_0 \hat{\vb z}$. In the absence of a large-scale electric field, the particle's motion is described by the Lorentz force:
\begin{equation}
\frac{d \vb p}{dt} = \frac{q}{c} (\vb v \wedge \vb B_0)
\label{eq:lorentz}
\end{equation}

The Lorentz force acts perpendicular to the particle's motion, preserving the velocity's magnitude. By splitting the motion into its components, we obtain:
\begin{equation}
m \gamma \frac{d \vb v}{dt} = \frac{q}{c} (\vb v \wedge \vb B_0) \quad \rightarrow \quad 
\begin{cases}
m \gamma \frac{dv_x}{dt} = \frac{q}{c} v_y B_0 \\
m \gamma \frac{dv_y}{dt} = -\frac{q}{c} v_x B_0\\
\frac{dv_z}{dt} = 0
\end{cases}
\label{eq:lorentzcomponents}
\end{equation}

The last equation shows that $v_z = v_\| = v \cos \theta$ remains constant. Consequently, the pitch angle, defined as the cosine of the angle between the particle velocity and the magnetic field direction ($\mu = \cos \theta$), is a conserved quantity, as $dv_z/dt = v d\mu/dt = 0$.

Combining the first two equations yields two second-order differential equations:
\begin{equation}
\frac{d^2 v_{x,y}}{dt^2} = - \Omega^2 v_{x,y}
\end{equation}

Here, we introduce the Larmor frequency $\Omega = \frac{q B_0}{m \gamma c}$.

This equation can be easily solved as simple harmonic motion along the $\hat{\vb x}$ axis, where $v_{x} = v_{0,x} \cos (\Omega t)$. Using this solution, we can solve the system~\eqref{eq:lorentzcomponents} as follows:
\begin{equation}
\begin{cases}
v_x = v_{0,\perp} \cos (\phi - \Omega t) \\
v_y = - v_{0,\perp} \sin (\phi - \Omega t) \\
v_z = v_{0,\parallel}
\end{cases}
\rightarrow \, 
\begin{cases}
v_x = v_{0} (1 - \mu^2)^{\frac{1}{2}} \cos (\phi - \Omega t) \\
v_y = - v_{0} (1 - \mu^2)^{\frac{1}{2}} \sin (\phi - \Omega t) \\
v_z = v_{0} \mu
\end{cases}
\end{equation}

Here, $\phi$ is an arbitrary phase, $v_{0,\perp}$ represents the initial velocity of the particle in the $xy$-plane, given by $v_{0,\perp} = v_0 \sin \theta = v_0 (1-\mu^2)^{1/2}$.

The solution above represents a helical motion with a uniform drift along $\hat{\vb z}$, described by the equation of motion $z = v \mu t$.

Now we consider introducing a perturbation to the magnetic field with components $\delta \mathbf{B} \equiv (\delta {\rm B}_x, \delta {\rm B}_y, \delta {\rm B}_z)$, where $|\delta \mathbf{B}| \ll |\mathbf{B}_0|$. In this case, we assume a pure Alfvénic wave propagating along the background magnetic field, which implies $\delta {\rm B}_z = 0$ and the wave oscillates such that $\delta \mathbf{B} \perp \mathbf{k}$.

This allows us to express the system of equations~\eqref{eq:lorentz} as follows:
\begin{equation}
m \gamma \frac{d\vb v}{dt} = \frac{q}{c}
\left|
\begin{array}{ccc}
\hat {\vb x}  & \hat {\vb y}  & \hat {\vb z}  \\
v_x & v_y & v_z \\
\delta {\rm B}_x & \delta {\rm B}_y & {\rm B}_0 
\end{array}
\right|
\oset{\delta {\rm B} \ll {\rm B}_0} \simeq
\frac{q}{c}
\left(
\begin{array}{c}
v_y {\rm B}_0 \\
-v_x {\rm B}_0 \\
v_x \delta {\rm B}_y - v_y \delta {\rm B}_x
\end{array}
\right)
\end{equation}

As prescribed by QLT, we neglect the perturbation field in the $x$ and $y$ components. This implies that the circular orbits in the plane perpendicular to the background field are approximately unaffected. However, the perturbation does cause a change in the $z$ component of the velocity, leading to a modification in the pitch angle $\mu$ of the particle. It's important to note that the perturbation does not affect the particle's momentum value; we are describing the motion in the reference frame of the perturbation, where the only force acting on the particle is the Lorentz force. Consequently, while numerous pitch-angle changes can eventually reverse the parallel velocity of the particle, they cannot shift the guiding center of the orbits.

To examine the extent of this change, we focus on the last equation of the system mentioned above, which governs the \emph{perturbed} motion along $z$:
\begin{equation}
m \gamma \frac{dv_z}{dt} = \frac{q}{c} \left[v_x(t) \delta {\rm B}_y - v_y(t) \delta {\rm B}_x \right]
\end{equation}

As a consequence, the pitch angle changes with time according to:
\begin{equation}
m\gamma v \frac{d\mu}{dt} = \frac{q}{c}v_{0,\perp} \left[\cos(\phi-\Omega t) \delta {\rm B}_y - \sin(\phi - \Omega t) \delta {\rm B}_x\right]
\label{eq:pitchanglemotion}
\end{equation}

To proceed, we make the simplifying assumption that the perturbed field is circularly polarized, meaning the wave components have the same amplitude: $|\delta {\rm B}_{x}| = |\delta {\rm B}_{y}| = |\delta {\rm B}|$. Thus, we can express the perturbation as:
\begin{equation}
\begin{cases}
\delta {\rm B}_y = &  \delta {\rm B} \exp \left[ i (kz - \omega t) \right] \\ 
\delta {\rm B}_x = & \pm i \delta {\rm B}
\end{cases}
\end{equation}

Taking the real part gives:
\begin{equation}
\begin{cases}\delta {\rm B}_y = & \delta {\rm B} \cos (kz - \omega t) \\
\delta {\rm B}_x = & \mp\delta {\rm B} \sin (kz - \omega t) 
\end{cases}
\end{equation}
therefore, by substituting in equation~\eqref{eq:pitchanglemotion}, we find
\begin{equation}
m\gamma v \frac{d\mu}{dt} = 
\frac{q}{c}v_{0,\perp} \delta{\rm B} \left[\cos(\phi-\Omega t) \cos (kz - \omega t) \pm \sin(\phi - \Omega t) \sin (kz - \omega t)\right]
\end{equation}
which simplifies to\footnote{We use the trigonometric relation $\cos \alpha \cos \beta \pm \sin \alpha \sin \beta = \cos (\alpha \mp \beta)$}:
\begin{equation}
m\gamma v \frac{d\mu}{dt} = 
\frac{q}{c}v_{0,\perp} \delta {\rm B} \cos(\phi-\Omega t \mp kz \pm \omega t)
\end{equation}

For Alfvén waves, the dispersion relation is given by $\omega = k v_{\rm A}$, where $v_{\rm A}$ represents the Alfvén velocity. By comparing the spatial frequency with the temporal frequency in the argument of the cosine function, we can derive the following relation:
\begin{equation}
\frac{kz}{\omega t} \simeq \frac{k v \mu t}{k v_{\rm A} t} \sim \frac{v}{v_{\rm A}} \mu
\label{eq:kmomegat}
\end{equation}

Here, we utilize the fact that for an unperturbed orbit, the position $z$ of the particle is given by $z = v \mu t$.

Considering that $v$ is of the order of the speed of light, while $v_{\rm A}$ in the average ISM is approximately 10 km/s, we find that the ratio in equation~\eqref{eq:kmomegat} is significantly greater than 1 unless $\mu \ll v_{\rm A} / v$.

Consequently, we can neglect the term $\omega t$ in comparison to $kz$. This choice is equivalent to selecting a reference frame in which the waves appear stationary. In this frame, there is no electric field associated with the waves.

We can approximate the pitch angle equation of motion as
\begin{equation}
\frac{d\mu}{dt} \simeq \Omega
(1-\mu^2)^{\frac 1 2} \frac{\delta {\rm B}}{{\rm B}_0} \cos\left[\phi + (\Omega \pm k v \mu) t \right]
\end{equation}

The equation above implies a periodic variation in the pitch angle. When we integrate this equation over a sufficiently long time interval, the average of the integrated quantity becomes zero. This result is physically expected since the particle orbits are concentric circles.

However, if we instead consider the square of the pitch-angle variation: 
\begin{multline}
\langle \Delta \mu \Delta \mu \rangle = \int_0^{2\pi} \frac{d\phi}{2\pi} \int_0^{\Delta t} dt \frac{d\mu}{dt}(t) \, \int_0^{\Delta t} dt^\prime \frac{d\mu}{dt}(t^\prime) \\ = \Omega^2 (1 - \mu^2) \left( \frac{\delta {\rm B}}{{\rm B}_0} \right)^2 \int_{0}^{\Delta t} dt \int_{0}^{\Delta t} dt^\prime \, \cos[(\Omega \pm k v \mu) t] \cos[(\Omega \pm k v \mu) t^\prime]
\end{multline}

The integrand functions are even, so we can double the interval of the $dt^\prime$ integral as $\int_{-\Delta t}^{\Delta t} dt^\prime$ and add a factor of $\frac{1}{2}$. Additionally, as we are considering sufficiently large times to evaluate the effect of scattering ($\Delta t \gg t, t^\prime$), the same interval can be approximated as $\int_{-\infty}^{\infty} dt^\prime$.

Therefore, we have\cite{Shalchi2009book}:
\begin{multline}
\langle \Delta \mu \Delta \mu \rangle = \\ \Omega^2 \frac{(1 - \mu^2)}{2} \left( \frac{\delta {\rm B}}{{\rm B}_0} \right)^2 \int_{0}^{\Delta t} \!\! dt \, {\rm Re} \{ \exp[i(\Omega \pm k v \mu) t] \} \, \int_{-\infty}^{\infty} \!\! dt^\prime \,  {\rm Re}\{\exp[i(\Omega \pm k v \mu) t^\prime]\}
\end{multline}
and solve the integral on $t^\prime$\footnote{We use the property $\delta(x-a) = \frac{1}{2\pi}\int_{-\infty}^\infty dy {\rm e}^{iy(x-a)}$}, as to obtain:
\begin{equation}
\langle \Delta \mu \Delta \mu \rangle = \Omega^2 \frac{(1 - \mu^2)}{2} \left( \frac{\delta {\rm B}}{{\rm B}_0} \right)^2 \int_{0}^{\Delta t} \!\! dt \, {\rm Re} \{ \exp[i(\Omega \pm k v \mu) t] \} \, 2\pi \delta (\Omega \pm k v \mu)
\end{equation}

The motion of the pitch angle, having null mean and variance different from zero, is typical of a diffusive process with diffusion coefficient $D_{\mu\mu}$, which represents the average rate of change of the square of the pitch angle over the time interval $\Delta t$.

Now the second integral, because of the presence of the~\emph{delta} function, gives just a factor $\int_{0}^{\Delta t} dt = \Delta t$, and we can write:
\begin{equation}
D_{\mu\mu} \equiv \left\langle \frac{\Delta \mu \Delta \mu}{\Delta t} \right\rangle = \Omega^2 \left( \frac{\delta {\rm B}}{{\rm B}_0} \right)^2 (1 - \mu^2) \, \pi \delta(\Omega \pm k v_{\parallel})
\label{eq:dmumuvar}
\end{equation}

In general, one must consider a packet of turbulent waves with energy distribution per wave number denoted as $W(k) dk$. This distribution represents the energy density contained within the range of wavenumbers $[k, k + dk]$ and is normalized to the energy density of the background magnetic field, $\frac{B_0^2}{8\pi}$. Specifically:
\begin{equation}
\left( \frac{\delta {\rm B}(k)}{{\rm B}_0} \right)^2 = W(k) dk.
\end{equation}

By incorporating this consideration, we can extend equation~\eqref{eq:dmumuvar} to obtain:
\begin{equation}
D_{\mu \mu} \equiv \left\langle \frac{\Delta \mu \Delta \mu}{\Delta t} \right\rangle = \Omega^2 (1 - \mu^2) \pi \int dk \, W(k) \delta(\Omega \pm k v_{\parallel}) \, .
\end{equation}

Introducing the resonant wavenumber $k_{\rm res}$, defined as the inverse of the Larmor radius $k_{\rm res} = r_{\rm L}^{-1} = \Omega/ v_\|$, we can express $D_{\mu \mu}$ as follows\footnote{We use the property $\int dx \delta (c x) = \frac{1}{|c|} \int dx \delta (x)$}:
\begin{equation} 
D_{\mu \mu} = \Omega (1 - \mu^2) \pi k_{\mathrm{res}} \int dk \, W(k) \delta(k \pm k_{\mathrm{res}}) = \Omega (1 - \mu^2) \pi k_{\mathrm{res}} W(k_{\rm res})
\end{equation}

These equations reveal that a wave-particle interaction is only possible when the inverse Larmor radius of the particle matches (i.e., is resonant) with the wavenumber of the turbulent wave. This type of process is commonly referred to as \emph{gyroresonant} scattering\footnote{It is worth noting that the QLT is consistently inadequate when attempting to describe pitch-angle diffusion at 90 degrees ($\mu = 0$) and reversing direction becomes a consideration. To address these and other limitations, several Nonlinear Theories have been formulated and developed.}.

The typical diffusion time, defined as the timescale to invert the pitch angle by about one radian is
\begin{equation}
\tau_{\rm diff} \simeq \frac{1}{D_{\theta\theta}} = \frac{1-\mu^2}{D_{\mu\mu}} = \frac{1}{\pi \Omega k_{\rm res} W(k_{\rm res})}
\end{equation}
where $D_{\theta\theta}$ is the diffusion coefficient in angle.

As in a diffusion timescale the particle moves by a distance of about $\Delta z = v \tau_{\rm diff}$, the spatial diffusion coefficient coefficient can be roughly estimated as
\begin{equation}
D_{zz} = v (v \tau_{\rm diff}) = \frac{v^2}{\pi \Omega k_{\rm res} W(k_{\rm res})} \simeq \frac{1}{3} r_{\rm L} v \frac{1}{k_{\rm res} W(k_{\rm res})} 
\end{equation}
where we have made the approximation $\pi \sim 3$. This informs us that the spatial diffusion coefficient is always much larger than the Bohm diffusion ($D_{\rm B} = \frac{1}{3} r_{\rm L} v$) since $k_{\rm res} W(k_{\rm res}) \ll 1$ at the relevant scales.
\section{The weighted slab technique}
\label{app:wsbla}

At present, the diffusion-losses model with the potential inclusion of advection is considered as the most suitable description for CR transport in the Galaxy at energies below approximately $10^{15}$ eV. Therefore, in this appendix we present the generalization of the transport model described in section~\ref{sec:nuclei} by including advection and ionization energy losses. 
This model is referred to as the \emph{weighted slab model}~\cite{Ptuskin1996apj,Jones2001apj}.

The distribution function $f_\alpha(z, p)$, which characterizes a stable species $\alpha$ subject to spallation and ionization losses, follows the transport equation:
\begin{multline}
-\frac{\partial}{\partial z} \left[ D_{\alpha}(p) \frac{\partial f_\alpha}{\partial z} \right]
+ u_z(z) \frac{\partial f_\alpha}{\partial z}
- \frac{du_z}{dz} \frac{p}{3} \frac{\partial f_\alpha}{\partial z}
+ \frac{f_\alpha}{\tau_{\rm sp,\alpha}}
+ \frac{1}{p^2} \frac{\partial}{\partial p}(p^2 \dot{p}_\alpha f_\alpha ) \\
= Q_\alpha(p,z) + \sum_{\alpha'>\alpha} \frac{f_{\alpha'}}{\tau_{\rm sp, \alpha'}}
\label{eq:fulltransport}
\end{multline}

Here, $D_\alpha(p, z)$ represents the diffusion coefficient of the particle, $u(z)$ is the advection velocity along the $z$-direction, and the term proportional to $du_z/dz$ accounts for adiabatic energy losses. The timescale for spallation is denoted as $\tau_{\rm sp}$, and $\dot{p} = dp/dt < 0$ captures ionization losses. The right-hand side of the equation includes the primary source term $Q_\alpha(p,z)$ and the contribution from the fragmentation of heavier nuclei.

Due to the coupling induced by nuclear fragmentation, to model CR propagation, it is required to solve a set of approximately $\sim$~100 coupled transport equations for all isotopes involved in the nuclear chain, accounting for all nuclear fragmentation yields.

In the one-dimensional slab model, a thin disk of matter represents the CR sources, $Q_\alpha(p,z) = 2 h \delta(z) Q_{\alpha, 0}(p)$, while an extended halo corresponds to the region where CRs diffuse with a diffusion coefficient that depends on rigidity.
In this model, we also make the assumption that $D$ and $u$ remain constant in the halo, and the advection velocity changes sign above and below the plane. Specifically, $u(z) = u_0 \left[2\Theta(z)-1\right]$, and consequently its derivative reads $\frac{du}{dz} = 2 u_0 \delta(z)$.

The majority of matter concentrates within the disk, characterized by a density $n_{\rm d}$, while the halo's density, $n_{\rm H}$, is considered negligible. Therefore, spallation and ionization losses only occur within the thin disk, resulting in $\dot{p}_\alpha(p,z) = 2 h \delta(z) \dot{p}_{0,\alpha}(p)$. It is important to note that this approximation is valid when $n_{\rm H} H \lesssim n_{\rm d} h$. If this condition is not satisfied, spallation in the halo can no longer be neglected.

To incorporate the presence of helium as a target in the ISM into the spallation term, we write the term for total inelastic fragmentation as:
\begin{equation}
\frac{f}{\tau_{\rm sp}} = v [n_{\rm H} \sigma_{\rm H} + n_{\rm He} \sigma_{\rm He}] f = v n_{\rm H} \sigma_{\rm H} [1 + f_{\rm He} \Sigma_{\rm He}] f
\end{equation}

Here, $f_{\rm He} = n_{\rm He} / n_{\rm H} \simeq 0.08$ represents the number density fraction of helium in the ISM, and $\Sigma_{\rm He} = \sigma_{\rm He} / \sigma_{\rm H}$ is the ratio between the inelastic cross-sections on helium and on hydrogen (with $\sigma_{\rm H}$ and $\sigma_{\rm He}$ being the respective cross-sections).

By expressing the interstellar gas density as:
\begin{equation}
\rho_{\rm ISM} = m_p n_{\rm H} + m_{\rm He} n_{\rm He} = m_p n_{\rm H} [ 1 + 4 f_{\rm He}]
\end{equation}
we obtain:
\begin{equation}
\frac{f}{\tau_{\rm sp}} = v \rho \frac{\sigma_{\rm H}}{m_p} \frac{[1 + f_{\rm He} \Sigma_{\rm He}]}{[1 + 4 f_{\rm He}]} f = \rho \frac{v \sigma_\alpha}{m} f = \mu \delta(z) \frac{v \sigma_\alpha}{m} f
\end{equation}
where $\mu = 2.3$ g cm$^{-2}$ represents the ISM surface density, and we define:
\begin{equation}
\sigma_\alpha \equiv \sigma_{\rm H} \frac{1 + f_{\rm He} \Sigma_{\rm He}}{1 + f_{\rm He}} \, , \,\,\, m \equiv m_p \frac{1 + 4 f_{\rm He}}{1 + f_{\rm He}} 
\end{equation}

Adopting the assumptions described earlier, we can derive the transport equation as follows.
The transport equation in \ref{eq:fulltransport} takes the form:
\begin{multline}
-\frac{\partial}{\partial z} \left[ D_{\alpha} \frac{\partial f_\alpha}{\partial z} \right]
+ u_0 [2 \Theta(z) - 1] \frac{\partial f_\alpha}{\partial z}
- \frac{2}{3} u_0 \delta(z) p \frac{\partial f_\alpha}{\partial p}
+ \frac{\mu v \sigma_\alpha}{m} \delta(z) f_\alpha
+ \frac{2 h}{p^2} \frac{\partial}{\partial p}(p^2 \dot{p}_0 f_\alpha ) \delta(z) \\
= 2h q_{0,\alpha} \delta(z) 
+ \sum_{\alpha'>\alpha} \frac{\mu v \sigma_{\alpha'\rightarrow\alpha}}{m} \delta(z) f_{\alpha'}
\label{eq:fulltransport2}
\end{multline}

To find a solution for this equation, we focus on the region outside the disc ($z \ne 0$), where the equation can be simplified as:
\begin{equation}
-D_{\alpha} \frac{\partial^2 f_\alpha}{\partial z^2} 
+ u_0 \frac{\partial f_\alpha}{\partial z} = 0
\end{equation}

We assume the solution takes the form $f = A {\rm e}^{r_+ z} + B {\rm e}^{r_- z}$, where $r_\pm$ are the roots of the equation $-D_{\alpha} r^2 + u_0 r = 0$, and we find that $r_+ = 0$ and $r_- = u_0 / D_\alpha$.

Using the boundary conditions $f(z=0) = f_0$ and $f(z=H) = 0$, we determine the coefficients $A$ and $B$ as:
\begin{equation}
A = f_0 \frac{{\rm e}^\xi}{{\rm e}^\xi - 1} \,\,\, , \,\,\, B = f_0 \frac{1}{1-{\rm e}^\xi}
\end{equation}

Here, $\xi = \frac{u_0 H}{D_\alpha}$, and the spatial distribution of the solution in the halo becomes:
\begin{equation}
f(z) = f_0 \frac{1-{\exp}\left[-\xi(1-\frac{|z|}{H})\right]}{1- \exp(-\xi)}
\end{equation}
where $z$ is taken with the absolute value as the solution is symmetric above and below the disc.

The derivative of $f_\alpha$ evaluated at the disc location can be expressed as:
\begin{equation}
\left.\frac{\partial f}{\partial z}\right|_0 = -\frac{f_0}{H} \frac{\xi}{\exp (\xi) - 1}
\label{eq:fgradient}
\end{equation}
which is proportional to the average gradient $\sim f_0/H$.

Next, we integrate equation~\ref{eq:fulltransport2} in an infinitesimal height around the disc, resulting in:
\begin{multline}
-2 D_{\alpha} \frac{\partial f_{0,\alpha}}{\partial z} 
- \frac{2}{3} u_0 p \frac{\partial f_{0,\alpha}}{\partial p}
+ \frac{\mu v \sigma_\alpha}{m} f_{0,\alpha}
+ \frac{2 h}{p^2} \frac{\partial}{\partial p}(p^2 \dot{p}_{0,\alpha} f_{0,\alpha} ) \\
= 2h q_{0,\alpha} 
+ \sum_{\alpha'>\alpha} \frac{\mu v \sigma_{\alpha'\rightarrow\alpha}}{m}  f_{0,\alpha'}
\label{eq:fulltransportabovebelow}
\end{multline}

Here the symmetry of the problem above and below the disc, namely $f_\alpha(z, p) = f_\alpha(-z, p)$, leads to the factor of two in front of the diffusion term and lets the advection term disappear.

Using the spatial derivative of $f_\alpha$ (equation~\ref{eq:fgradient}) reduces the equation to a first-order differential equation for $f_{0,\alpha}(p)$:
\begin{multline}
\frac{2 u_0}{\mu v} \frac{1}{{\rm e}^\xi - 1} f_{0,\alpha}
- \frac{2 u_0}{3 \mu v}  p \frac{\partial f_{0,\alpha}}{\partial p}
+ \frac{2 h}{\mu v p^2} \frac{\partial}{\partial p}(p^2 \dot{p}_{0,\alpha} f_{0,\alpha} ) 
+ \frac{\sigma_\alpha}{m} f_{0,\alpha} \\
= \frac{2h}{\mu v} q_{0,\alpha} 
+ \sum_{\alpha'>\alpha} \frac{\sigma_{\alpha'\rightarrow\alpha}}{m}  f_{0,\alpha'}
\end{multline}

Multiplying both sides by $A p^2$ we express the equation in terms of the intensity as a function of the kinetic energy per nucleon $E$, which relates to $f_\alpha$ as described in equation~\ref{eq:f2I}: 
\begin{multline}
\frac{2 u_0}{\mu v} \frac{1}{{\rm e}^\xi - 1} I_\alpha
- \frac{2 u_0}{3 \mu v} p^3 \frac{\partial}{\partial p} \frac{I_\alpha}{p^2}
+ \frac{2 h}{\mu v} \frac{\partial}{\partial p}(\dot{p}_0 I_\alpha) 
+ \frac{\sigma_\alpha}{m} I_\alpha \\
= \frac{2h A p^2}{\mu v} q_{0,\alpha} 
+ \sum_{\alpha'>\alpha} \frac{\sigma_{\alpha'\rightarrow\alpha}}{m}  I_{\alpha'}
\end{multline}
where $A$ is the nuclear mass number.

To simplify the equation, we add and subtract the term $\frac{2u_0}{\mu v} I_\alpha$, resulting in:
\begin{multline}
\frac{2 u_0}{\mu v} \frac{1}{1-{\rm e}^{-\xi}} I_\alpha
- \frac{2 u_0}{3 \mu v} I_\alpha 
- \frac{2 u_0}{3 \mu v} p \frac{\partial I_\alpha}{\partial p} 
+ \frac{2 h}{\mu v} \frac{\partial}{\partial p}(\dot{p}_{0,\alpha} I_\alpha) 
+ \frac{\sigma_\alpha}{m} I_\alpha \\
= \frac{2h A p^2}{\mu v} q_{0,\alpha} 
+ \sum_{\alpha'>\alpha} \frac{\sigma_{\alpha'\rightarrow\alpha}}{m}  I_{\alpha'}
\end{multline}

This equation can be rewritten conveniently as:
\begin{multline}
\frac{I_\alpha(E)}{\rchi_\alpha}
+ \frac{d}{dE} \left( \left[ \left( \frac{dE}{d\rchi}\right)_{\rm ad} + \left( \frac{dE}{d\rchi} \right)_{\rm ion, \alpha} \right] I_\alpha(E) \right)
+ \frac{I_\alpha(E)}{\rchi_{\rm cr, \alpha}} \\
= \frac{2h A p^2}{\mu v} q_{0,\alpha} 
+ \sum_{\alpha'>\alpha} \frac{\sigma_{\alpha'\rightarrow\alpha}}{m}  I_{\alpha'}
\label{eq:fulltransport3}
\end{multline}

Here, we introduce the effective grammage:
\begin{equation}
\rchi_\alpha = \frac{\mu v}{2 u_0} \left[ 1 - \exp \left(-\frac{u_0 H}{D_\alpha} \right) \right]
\end{equation}
and the term\footnote{We use $\partial p/\partial E = A / c \beta$}:
\begin{equation}
 \left( \frac{dE}{d\rchi}\right)_{\rm ad} + \left( \frac{dE}{d\rchi} \right)_{\rm ion, \alpha} = 
 -\frac{2 u_0}{3 \mu c} \sqrt{E^2 + 2 m_p c^2 T} - \frac{2 h}{\mu A} |\dot p_{0,\alpha}|
\end{equation}

Notice that in the limit of $H^2 / D_\alpha \ll H / u_0$, the grammage reduces to the expression derived in the diffusion scenario in equation~\ref{eq:grammage}.

We observe that equation~\ref{eq:fulltransport3} can be written in the form:
\begin{equation}
\Lambda_{1,\alpha} I_\alpha(E) + \Lambda_{2, \alpha}(E) \partial_E I_\alpha(E) = Q_\alpha(E)
\end{equation}
which has a formal solution given by:
\begin{equation}
I_\alpha(E) = \int_E dE^\prime \frac{Q_\alpha(E^\prime)}{|\Lambda_{2,\alpha}(E^\prime)|} \exp\left[ -\int_E^{E^\prime} dE^{\prime\prime} \frac{\Lambda_{1,\alpha}(E^{\prime\prime})}{|\Lambda_{2,\alpha}(E^{\prime\prime})|} \right]
\end{equation}

In the more general case, this expression has to be solved numerically to derive the local interstellar spectrum for all isotopes in cosmic radiation\footnote{A practical application of this approach is implemented in the CRAMS code available at \url{https://github.com/carmeloevoli/crams}}.

Finally, we highlight that this formalism can be straightforwardly extended to model the transport of unstable nuclei, consider the effects of a finite thick disc $h$, or account for changes in the secondary production due to the grammage accumulated in the halo (see, e.g.,~\cite{Morlino2020prd,Evoli2020prd}).

\section{Green's functions of the transport equations}
\label{sec:appgreen}

\subsection{Green's function of the pure diffusive equation}

To construct the Green's function, we start by taking the Fourier transform with respect to the spatial variable $\vb r$, using the property that $\mathcal F[\delta^{(3)}(\vb r - \vb r_\star)] = {\rm e}^{-i \vb k \cdot \vb r_\star}$ :
\begin{equation}
\frac{\partial}{\partial t} \tilde{\mathcal G}(\vb k, t \leftarrow \vb r_\star, t_\star) 
+ D k^2 \tilde{\mathcal G}(\vb k, t \leftarrow \vb r_\star, t_\star) 
= {\rm e}^{-i \vb k \cdot \vb r_\star} \delta(t - t_\star)
\end{equation}
subject to the initial condition $\tilde{\mathcal G}(\vb k, 0 \leftarrow \vb r_\star, t_\star) = 0$.

After multiplying both sides by ${\rm e}^{D k^2 t}$, we obtain
\begin{equation}
\frac{\partial}{\partial t} \left[ {\rm e}^{D k^2 t} \tilde{\mathcal G}(\vb k, t \leftarrow \vb r_\star, t_\star) \right]
= {\rm e}^{-i \vb k \cdot \vb r_\star + D k^2 t} \, \delta(t - t_\star)
\end{equation}

This equation can be easily solved, leading to the Fourier transform of the Green's function:
\begin{equation}
\tilde{\mathcal G}(\vb k, t) = 
{\rm e}^{-i \vb k \cdot \vb r_\star - D k^2 t} 
\int_0^t {\rm e}^{D k^2 t^\prime} \delta(t^\prime-t_\star) dt^\prime
= \begin{cases}
0 & t < t_\star \\
{\rm e}^{- i \vb k \cdot \vb r_\star - D k^2 (t-t_\star)} & t > t_\star
\end{cases}
\end{equation}

The upper limit $t$ of this integral expresses \emph{causality}: the solution at time $t$ depends only on causes lying in its past, i.e., $t_\star < t$.

Finally, by taking the inverse Fourier transform with respect to $\vb k$, we obtain the Green's function we are looking for:
\begin{equation}
\mathcal G (\vb r, t \leftarrow \vb r_\star, t_\star) 
= \frac{\Theta(t-t_\star)}{(2\pi)^3} 
\int {\rm e}^{i \vb k \cdot (\vb r - \vb r_\star)} {\rm e}^{-Dk^2 (t-t_\star)} d^3 \vb k
\end{equation}

Recognizing the integral as the (inverse) Fourier transform of a Gaussian, we find:
\begin{equation} 
\mathcal G_{\rm free} (\vb r, t \leftarrow \vb r_\star, t_\star) 
= \frac{\Theta(\tau)}{(4\pi D \tau)^{3/2}} 
{\rm e}^{-\frac{d^2}{4 D \tau}}
\end{equation}
where $\tau = t-t_\star$ and $\vb d = \vb r - \vb r_\star$.

Thus, an initial Gaussian distribution retains its Gaussian form, with its squared width spreading linearly with time. This linear growth of variance is characteristic of \emph{diffusing} probabilistic processes.

One notable property of this solution is that $\mathcal G > 0$ everywhere for any finite $t > 0$, regardless of how small $t$ is. However, this violates Special Relativity, and it is often resolved by replacing the theta argument with $\Theta(c \Delta t - d)$.

It's worth mentioning that the halo size parameter $H$ does not appear in the Green's function, as we derived the free-space Green's function without imposing any boundary condition in the $z$ direction.

To enforce the desired boundary conditions, we introduce a set of image charges with coordinates~\cite{Cowsik1979apj,Baltz1998prd,Mertsch2011jcap}:
\begin{equation}
\vb r_{\star,n}^\prime = 
\left(\begin{array}{c}
x_\star\\
y_\star\\
2 H n + (-1)^n z_\star 
\end{array}\right)
\end{equation}
as a consequence, the Green's function associated with these image charges becomes
\begin{equation}
\mathcal G_{\rm H} (\vb r, t \leftarrow \vb r_\star, t_\star) = \sum_{n=-\infty}^{\infty} (-1)^n \mathcal G_{\rm free} (\vb r, t \leftarrow \vb r_{\star,n}^\prime, t_\star)  
\end{equation}

It is easy to check that $\mathcal G_{\rm H}(x, y, {z = \pm {\textrm H}}, t\leftarrow \vb r_\star, t_\star) = 0$, satisfying the desired boundary conditions.

\subsection{Green's function of the diffusion-losses equation}

The equation of interest for leptons, assuming steady-state, is given by:
\begin{equation}
- D(E) \nabla^2 n_e(E)
- \frac{\partial}{\partial E} \left[ b(E) n_e \right] 
= Q(\vb r, E, t)
\end{equation}
where $b(E) = dE/dt$ represents the energy losses.

It is convenient to introduce a new variable:
\begin{equation}
\tilde t = 4 \int_E^\infty \frac{D(E^\prime)}{b(E^\prime)} dE^\prime
\end{equation}
which allows us to rewrite the derivative as:
\begin{equation}
\frac{d}{d\tilde t} = - \frac{b(E)}{4 D(E)} \frac{d}{dE}
\end{equation}

Using this new variable, we obtain the equation:
\begin{equation}
- D(E) \nabla^2 n_e(E)
+ 4 \frac{D(E)}{b(E)}\frac{\partial}{\partial \tilde t} \left[ b(E) n_e(E) \right] 
= Q(\vb r, E, t)
\end{equation} 

By rearranging the terms, we arrive at the diffusion equation for the new variable $\tilde{n} = b(E) n(E)$:
\begin{equation}
\frac{\partial}{\partial \tilde t} \left[ \tilde n_e(E) \right] 
- \frac{1}{4} \nabla^2 \left[ \tilde n_e(E) \right]
= \tilde Q(\vb r, E, t)
\end{equation} 

Next, we consider the Green's function in the new variables, introducing $\lambda^2 = \tilde{t} - \tilde{t}_\star$, which is given by:
\begin{equation}
\tilde{\mathcal G} (\vb r, \tilde t \leftarrow \vb r_\star, \tilde t_\star) 
= \frac{\Theta(\lambda^2)}{(\pi \lambda^2)^{3/2}} 
{\rm e}^{-\frac{d^2}{\lambda^2}}
\end{equation}

The Green's function for the time-dependent solution can be expressed in terms of the steady-state solution as:
\begin{equation}
{\mathcal G} (\vb r, t, E \leftarrow \vb r_\star, t_\star, E_\star) 
= \delta(\Delta t - \tau) \frac{\tilde{\mathcal G} (\vb r, \tilde t \leftarrow \vb r_\star, \tilde t_\star)}{|b(E)|} 
= \frac{1}{|b(E)|}\frac{\delta(\Delta t - \tau)}{(\pi \lambda^2)^{3/2}} {\rm e}^{-\frac{d^2}{\lambda^2}}
\end{equation}
where $\lambda^2(E, E_\star) = 4 \int_{E}^{E_\star} dE^\prime \frac{D(E^\prime)}{b(E^\prime)}$ represents the propagation scale (also known as the Syrovatskii variable), and $\tau(E, E_\star) = \int_E^{E_\star} \frac{dE^\prime}{b(E^\prime)}$ is the loss time, which corresponds to the average time during which the energy of a particle decreases from $E_\star$ to $E$ due to losses.

Therefore, the particles we observe with energy $E$ have been injected with energy $\tilde{E}$ at a time $\Delta t$ that satisfies $\tau(E, \tilde{E}) = \Delta t$.
 
This condition sets a maximum energy $E_{\rm max}$ as $\tau(E_{\rm max}, \infty) = t_{\rm age}$, which in the Thomson limit, is given by:
\begin{equation}
E_{\rm max} = \frac{E_0^2}{b_0 t_{\rm age}} \simeq 400~{\rm GeV} \left( \frac{t_{\rm age}}{\rm Myr} \right)^{-1}
\end{equation}

This result provides the maximum energy of observed particles based on the age of the source.

\acknowledgments

We are grateful to P.~Blasi for providing the inspiration for these lecture notes, which stem from the course taught effortlessly over the past decade at GSSI.

We extend our sincere appreciation to our experimental colleagues, whose tireless efforts have consistently pushed the boundaries of knowledge and understanding in the subject matter. 

We would also like to acknowledge the usage of the CRDB~\cite{Maurin2023} database of CR measurements. 

Finally, CE would like to extend his heartfelt thanks to the Italian Physical Society for granting us the opportunity to host this School at the enchanting and cozy Villa Monastero in \emph{Varenna}.
%

\bibliography{2022-varenna.bib}
\bibliographystyle{varenna}

\end{document}